\documentclass[journal,onecolumn,12pt,twoside]{IEEEtranTCOM}
\normalsize 

\usepackage{mathtools}
\usepackage{amsmath}
\usepackage{amssymb}
\usepackage{amsmath}
\usepackage{cite}

\usepackage{graphicx}
\usepackage{graphicx}
\usepackage{mdwmath}
\usepackage{eqparbox}
\usepackage{fixltx2e}
\usepackage{stfloats}
\usepackage[section]{placeins}
\ifCLASSINFOpdf
\else
\fi

\usepackage{soul}

\usepackage{soulpos}

\begin{document}
%
\title{Statistical Modeling and Performance Characterization of Ultrafast Digital Lightwave Communication System Using Power-Cubic \mbox{Optical Nonlinear Preprocessor (Extended Version)}}
%
%
%

\author{Mahdi~Ranjbar~Zefreh,~\IEEEmembership{Student Member,~IEEE}~and~Jawad~A.~Salehi,~\IEEEmembership{Fellow,~IEEE}
\thanks{Corresponding author: J. A. Salehi; Professor of Electrical Engineering, Sharif University of Technology, jasalehi@sharif.edu.}
\thanks{Authors are with the Department
of Electrical Engineering, Sharif University of Technology, Tehran,
Iran e-mail: mahdi\_ranjbar@ee.sharif.edu, jasalehi@sharif.edu.}}
\maketitle

\begin{abstract}
\boldmath
In this paper, we present an analytical approach in obtaining the probability density function (pdf) of the random decision variable $Y$, formed at the output of power-cubic all-optical nonlinear preprocessor followed by the photodetector. Our approach can be used to accurately evaluate the performance of ultrafast pulse detection in the presence of Gaussian noise. Through rigorous Monte-Carlo simulation, the accuracy of widely used Gaussian approximation of decision variable $Y$ is refuted. However, in this paper we show that the so called Log-Pearson type-3 probability density function (LP3 pdf) is an excellent representation for the decision variable $Y$. Three distinguishable parameters of the LP3 pdf are obtained through analytical derivation of three moments of the decision variable $Y$. Furthermore, toward a more realistic model, in addition to ASE Gaussian noise, the effects of shot and thermal noises are also included. Finally, using the presented analytical approach, it is shown that power-cubic preprocessor outperforms its quadratic counterparts, i.e., Second Harmonic Generation (SHG) and Two Photon Absorption (TPA) devices, in high power regime where shot and thermal noises can be neglected. 
\end{abstract}


%
\IEEEpeerreviewmaketitle
\section{Introduction \label{sec_intro}}
%
%
%
%
\IEEEPARstart{D}{ue} to the ubiquitous use of ultrahigh bandwidth optical fiber medium, worldwide, and its potential for transmitting ultrafast digital lightwave information, optical fiber telecommunication system must be able to employ ultrashort light pulses with pico to femtosecond duration for future optical data communication networks \cite{lee2013microwave}. When data pulses, in the presence of noise, are communicated via ultrashort light pulses, the need to change the domain from high-speed all-optical signal processing to low-speed electrical-signal processing domain for ultrashort light pulse detection becomes an essential task. In practice, the source of the aforementioned noise, could be the amplified spontaneous emission (ASE) due to laser source and optical amplifiers in an optically-demultiplexed bit stream in an ultrafast OTDM system \cite{harstead2012future,mulvad20091,ji20101} or multiple-access interference (MAI) noise due to the coherent spectrally phase encoded optical code division multiple access (SPE-OCDMA) network, which is a promising candidate for future passive optical networks (PON) \cite{ghafouri2012optical,salehi2007emerging}. Ultrahigh speed photodetectors and ultrafast electronic circuitries with bandwidths on the order of terahertz, necessary for optimum detection of ultrashort light pulses, is believed to be unrealistic, complex and expensive. On the other, hand the direct detection of such optical pulses by band-limited conventional photodetectors would dramatically degrade the performance of such systems. However, a sub-optimum solution is contrived by placing nonlinear all-optical pre-processing device prior to the conventional band-limited photodetectors \cite{agrawal2010applications}. Many nonlinear all-optical preprocessors such as two-photon absorption (TPA) detectors, second-harmonic generation (SHG) crystals, nonlinear polarization rotation mirrors, highly birefringent photonic crystal fiber and spectral broadening fiber-based techniques have been proposed and successfully implemented in many experiments as alternative suboptimum approaches for the detection of ultrashort light pulses \cite{guo2009elimination,fsaifes2010nonlinear,jiang2005four,scott2005eight,chen2013effect,wang2005dispersion}. Kravtsov, Prucnal and Bubnov in \cite{kravtsov2007simple} proposed, and experimentally investigated, an ultrafast nonlinear thresholder based on a modified nonlinear optical loop mirror (NOLM). The key advantages of NOLM, such as its polarization insensitivity, multi-wavelength operational potential and simple all-fiber based structure, make this thresholder highly desirable. Furthermore, according to its cubic power transfer function, the systems utilizing a NOLM-based thresholder can potentially outperform systems utilizing SHG or TPA with quadratic power transfer function in the context of ultrashort light pulse detection. Indeed, higher order nonlinearity in the power-cubic device with respect to power-quadratic devices, eventuates a better discrimination between the high peak power ultrashort pulse and low average power noise \cite{farhang2009optimum}. Jamshidi and Salehi in \cite{jamshidi2006statistical,jamshidi2007performance} obtained a statistical model for TPA receivers based on Gaussian approximation of the decision variable. Also in \cite{ni2007performance} Ni, Lehnert and Weiner modeled SPM and SHG-based thresholder devices for use in SPE-OCDMA systems. Matinfar and Salehi in \cite{matinfar2009mathematical,matinfar2011performance} studied the SHG device as a preprocessor in ultrashort light pulse detection with Gaussian background noise and SPE-OCDMA systems with multiple-access noise. They succeeded in accomplishing in-depth statistical studies in order to model the SHG-based systems in thin and thick crystal regimes to evaluate the error probability of the proposed system. NOLM-based power-cubic thresholder is also theoretically investigated in \cite{kravtsov2009ultrashort}. While in the aforementioned article it has been emphasized that inclusion of beat noise to the model would be important, however for the sake of simplicity, the effect of beat noise is neglected. Also, in \cite{kravtsov2009ultrashort} the photodetector is modeled as a linear device with respect to the input optical power. 
\\
In this paper we intend to complete and fulfill an in-depth analysis and study on the statistical behavior of an ultrafast digital lightwave communication system based on a power-cubic nonlinear preprocessor. In addition to ASE noise that is modeled as an additive Gaussian noise due to optical amplifier, photodetector shot noise and electronic circuit thermal noise are included in our analytical studies. Beat noise, being a serious limiting case in many optical systems \cite{wang2004analysis,pu2006evaluation}, is also taken into account. Furthermore, as a more realistic model, the photodetector is modeled as an absolute squared energy detector ,with respect to input electrical field \cite{ni2007performance,matinfar2009mathematical,matinfar2011performance}. Toward mathematical realization, the LP3 probability distribution is presented as an excellent approximation for modeling the statistical behavior of the decision variable. The moments of the decision variable are obtained through elaborate statistical calculations. Since any appropriate LP3 distribution is uniquely distinguished by its three characteristic parameters, we have successfully determined our desirable LP3 pdf based on calculated moments of decision variable. The effects of shot and thermal noises are also included to the model, and finally, the system error probability expression is obtained and calculated numerically. 
\\
It must be noted that in the SPE-OCDMA system utilizing On-Off Keying (OOK) modulation, after the spectral phase decoding at the receiver, the ultrashort pulse must be efficiently detected despite Gaussian noise due to multiple access interference (MAI) and ASE. In the same fashion as \cite{jamshidi2006statistical,matinfar2009mathematical} which serves as pioneer works for \cite{jamshidi2007performance,matinfar2011performance}, the results presented in this paper would shed light on the performance studies of SPE-OCDMA receiver using power-cubic preprocessor.
\\ 
The rest of this paper is organized as follows. In section \ref{sec_system_math_model} the system description and its equivalent mathematical model are presented. In section \ref{sec_ver_DV}, an accurate model (i.e., LP3) for the probability density function of the decision variable is obtained. Section \ref{sec_ver_Dv_LP3} presents the method of finding the parameters of the probability density function based on the moments of the decision variable. In section \ref{sec_sh_th_effect}, effects of shot and thermal noises on the decision variable are discussed, and section \ref{sec_error_prob_and_opt_det}, discusses the methodology in obtaining the receiver's optimum threshold and error probability. In section \ref{sec_Num_Res}, the numerical results and error probability curves based on the analytical model obtained in previous sections are plotted and discussed. Section \ref{sec_conclusion} concludes the paper.

\section{System Description and Mathematical Model\label{sec_system_math_model}}
\begin{figure}[ht]
\centering
\includegraphics[scale=.75]{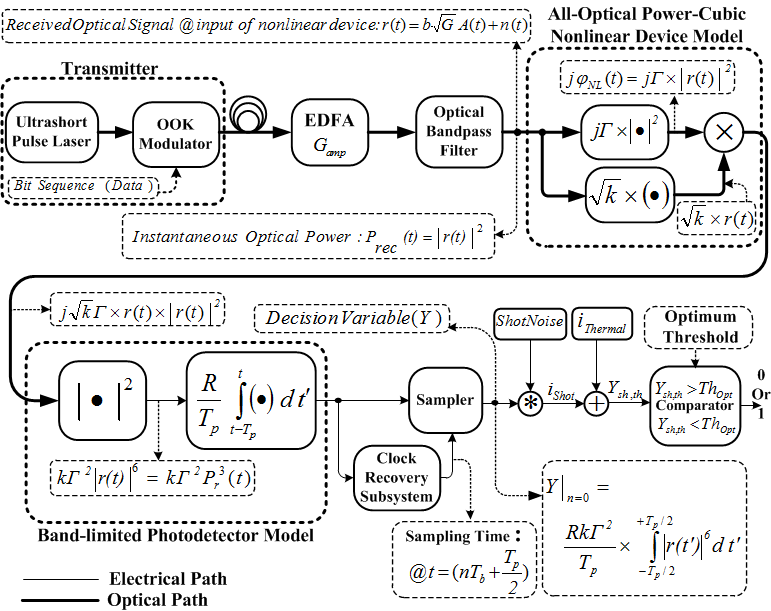}
\caption{transmitter-receiver model of ultrafast digital lightwave communication system based on power-cubic preprocessor.\label{fig_system_model}}
\end{figure}

A typical ultrafast digital lightwave communication system using a power-cubic optical nonlinear preprocessor is shown in \figurename{ \ref{fig_system_model}}. Assuming On-Off Keying (OOK) modulation at the transmitter, the receiver must have the ability to correctly detect bits, 0 or 1, sent by the transmitter. To this end, we first place an erbium-doped fiber amplifier (EDFA) for amplification of the received optical signal to repel the unfavorable effects of shot and thermal noises at the front end of the receiver. An optical band-pass filter eliminates out of band ASE noise. The output of the filter encounters the nonlinear power-cubic preprocessor followed by a photodetector. After clock recovery and sampling, the decision variable is compared with a threshold level in the comparator to determine the transmitted bit. According to the model presented in \cite{kravtsov2007simple}, in the case of absolute initial balance of the loop in nonlinear optical loop mirror (NOLM) and assuming small nonlinear shift regime ($\phi_{NL}=\Gamma{{P_r}^2}<<1$), the preprocessor acts as a cubic-law device. Hence, the output instantaneous power of the preprocessor is proportional to cubic power of input instantaneous optical power ($P_{out}=k{\Gamma}^2P_{in}^3(t)$) where $k$ and $\Gamma$ are two constants related to the physical structure of the NOLM device. 

Following the photodetector model as an energy detector, the photodetector obtains the average power of the input signal within its response time  \cite{matinfar2009mathematical,matinfar2011performance}. The mathematical expression for the sampled output electrical current of the photodetector ($Y$), neglecting effects of shot and thermal noises, is expressed as follows:
\begin{equation}
\label{DV main equation}
Y=\frac{Rk{\Gamma}^2} {T_p} \int^{+{T_p}/{2}}_{-{T_p}/{2}} {{\left|r\left(t^{'}\right)\right|}^6{dt}^{'}}
\end{equation}
Where, $T_p$ is the response time of the band-limited photodetector, inversely proportional to photodetector bandwidth. Also $R$ is the responsivity of the receiver's photodetector which is equal to $R=({{\eta}q_e})/({h\nu})$ \cite{govind2010fiber} where $\eta$ is the quantum efficiency of the photodetector, $q_e$ is the electron charge, $h$ is the Planck constant and $\nu$ is the incident light frequency. In \eqref{DV main equation}, $r\left(t\right)$ denotes the received optical signal after the blocks of EDFA and bandpass filter. Indeed, $r\left(t\right)$ is the envelope of the optical electrical field normalized such that the instantaneous optical power equals to the absolute square of the signal ($P_{rec}\left(t\right)={\left|r\left(t\right)\right|}^2$). Due to OOK modulation, $r\left(t\right)$ consists of an ultrashort pulse and noise if the transmitted bit is one, and noise only if the transmitted bit is zero. Following the notation of \cite{matinfar2009mathematical}:
\begin{equation}
\label{r_def}
{r\left(t\right)=b\sqrt{G}A\left(t\right)+n\left(t\right)=ba(t)+n(t)}
\end{equation}
 where $a(t)=\sqrt{G}A\left(t\right)$ and $b$ corresponds to the transmitted bit, 0 or 1, and $G$ is the total power gain experienced by the signal in the path from the transmitter laser source to the input of the preprocessor of the receiver, rephrased from \cite{jamshidi2005statistical} as; $G=G_{amp}L_1L_2$ in which $G_{amp}$ is EDFA gain, $L_1$ and $L_2$ are the loss before and after amplifier, respectively. Also $A(t)$ is the transmitted ultrashort optical light pulse and $n(t)$ is the filtered amplified spontaneous emission (ASE) due to EDFA. The output temporal pulse shape $A(t)$ of the light source, i.e., mode-locked laser (MLL), at the transmitter is considered to be a Sinc-function $A\left(t\right)=\sqrt{P_0}Sinc({t}/{\tau_c})$ in which $\tau_c$ and $P_0$ denote pulse duration and peak power, respectively \cite{ni2007performance, matinfar2009mathematical,salehi1990coherent}. At the input of the nonlinear device in the receiver, the pulse amplitude is scaled with the total power gain, i.e.,
 \begin{equation}
 \label{a_def}
 { a\left(t\right)\triangleq \sqrt{G}A\left(t\right)=\sqrt{P_r}\ Sinc\left({t}/{\tau_c}\right)}
 \end{equation}
 where $P_r\triangleq GP_0$ in \eqref{a_def}. The baseband equivalent of filtered ASE in the input of NOLM device can be expressed as $n\left(t\right)=p\left(t\right)+jq(t)$ which is assumed to be a complex band-limited white Gaussian process with $p\left(t\right)$ and $q\left(t\right)$ as its quadrature components. Indeed, $p\left(t\right)$ and $q\left(t\right)$ are Gaussian independent stationary processes with zero mean and autocorrelation functions as:
 \begin{equation}
 \label{atocorr_func_noise}
 {R_p\left(\tau\right)=R_q\left(\tau\right)=\sigma_0^2\ Sinc({\tau}/{{\tau}_c})}
 \end{equation}
 where $\sigma_0^2$  can be expressed as $\sigma^2_0=({\delta{L_2}})/({2{\tau}_c})$. In this expression $L_2$ and ${\delta}/{2}$ denote the loss after the amplifier and the two-sided power spectral density of ASE noise, respectively. Furthermore $\delta\triangleq n_{sp}\left(G_{amp}-1\right)h\nu$ \cite{humblet1991bit} where $n_{sp}$ is the spontaneous parameter of EDFA. An important parameter is yet to be defined is the processing ratio of the photodetector, i.e., $PRD={T_p}/{{\tau}_c}$, which will be used extensively in the next sections \cite{matinfar2009mathematical,matinfar2011performance}. Using the model presented in this section, a step-by-step analysis of the system performance is carried out in the following.

\section{Verifying the Decision Variable ($Y$) Probability Density Function Using Monte-Carlo Simulations \label{sec_ver_DV}}
\begin{figure}[ht]
\centering
\includegraphics[scale=.8]{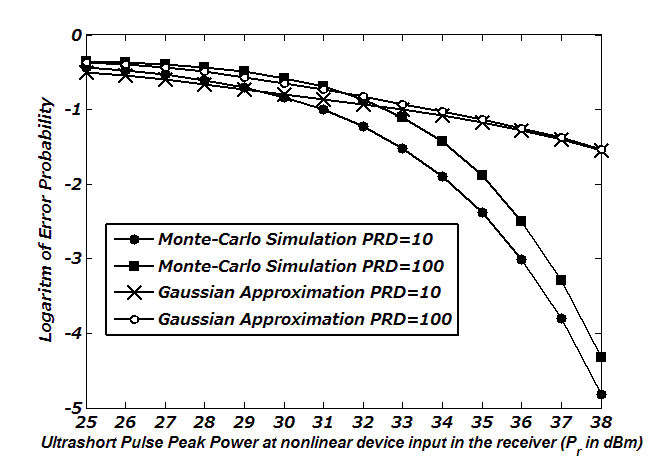}
\caption{Error probability based on Monte-Carlo approach and Gaussian approximation comparison.\label{fig_PE_m-c_Gauss_comp}}
\end{figure}

\begin{figure}[h]
\centering
\includegraphics[scale=.55]{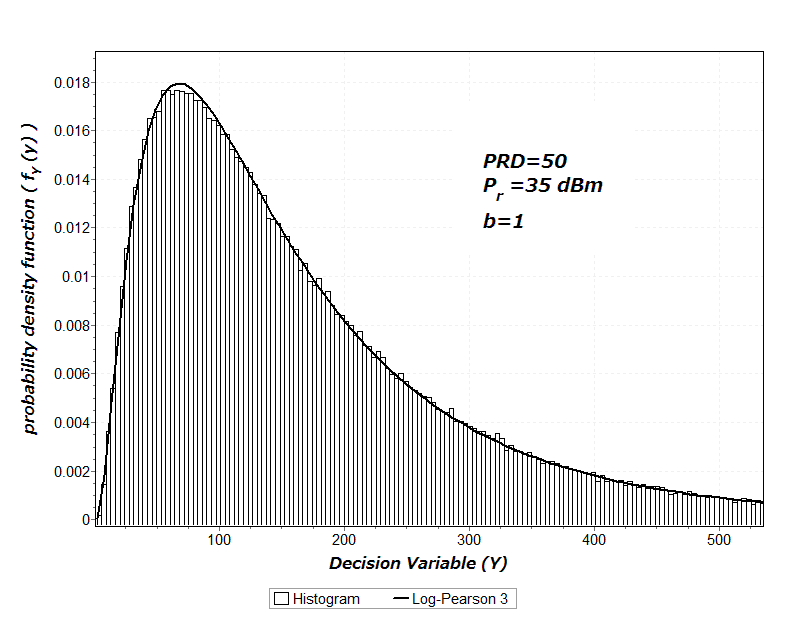}
\caption{Histogram due to transmitted bit one and fitted LP3 pdf.\label{fig_b_1_hist_comp_LP3}}
\end{figure}

\begin{figure}[h]
\centering
\includegraphics[scale=.55]{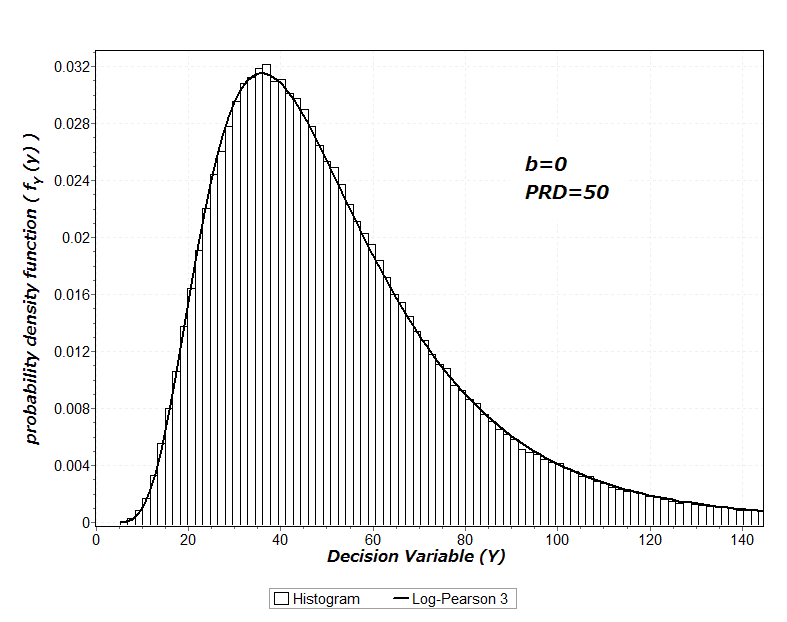}
\caption{Histogram due to transmitted bit zero and fitted LP3 pdf.\label{fig_b_0_hist_comp_LP3}}
\end{figure}

In this  section the statistical distribution of the decision variable $Y$, expressed in equation \eqref{DV main equation}, is investigated. We only consider ASE noise and shot and thermal noises are neglected in this section whereas their effects are taken into account in section \ref{sec_sh_th_effect}. The values of parameters used in simulations in this section are the same as those specified in \tablename{ \ref{table_numeric_values}}, unless mentioned otherwise.
\begin{table}[ht]
\caption{Numerical value of parameters used for calculations} 
\centering 
\begin{tabular}{c c c} 
\hline\hline 
Parameter &	Symbol &	Value \\ [0.5ex]
\hline %
Ultrashort Pulse width &	$\tau_c$ &	$100 fsec$ \\
Processing Ratio of detector &	$PRD$	& $10,25,50,100$ \\
Optical Wavelength &	$\lambda$	& $1.55 \mu m$ \\
Gain of EDFA &	$G_{amp}$ &	$50 dB$ \\
Receiver temperature &	$T_r$ &	$300 K$ \\
Input-Output Gain of NOLM based Preprocessor &	$k$ &  	$0.01$ \\
Nonlinear phase coefficient of NOLM &	$\Gamma$  &	$0.1 W^{-1}$ \\
Path Loss after EDFA &	$L_2$ &	$0 dB$ \\
Electrical Circuit Equivalent Load &	$R_L$ &	$100\Omega$ , $1 K\Omega$, $10 K\Omega$ \\
Spontaneous Emission Coefficient of EDFA &	$n_{sp}$ &	$1.1$ \\
Quantum Efficiency of Photodetector &	$ \eta $ &	$0.8$ \\
Dielectic Constant of SHG Crystal used for comparison &	$\kappa$ &	$2$ \\
Second Order Susceptibility of SHG & $\chi$ &	$2.1 \times 10^{-3}$ \\
Group Velocity Mismatch of SHG crystal &	$GVM$ &	$3\times 10^{-10} s/m$ \\
Response Time of Linear Filter in SHG Crystal Model\cite{matinfar2011performance} &	$T_L$ &	$0.1\times \tau_c=10 fsec$ \\
[1ex]

\hline 
\end{tabular}
\label{table_numeric_values} 
\end{table}
Precise evaluation of the system performance requires obtaining the probability density function (pdf) of the decision variable ($Y$). As a rule of thumb, Gaussian assumption of the decision variable has been held for many applications \cite{jamshidi2006statistical,jamshidi2007performance,kravtsov2009ultrashort,ni2007performance}. Hence, the Gaussian approximation has been employed for power-linear receiver bit error rate analysis. The decision variable in the case of the power-linear receiver could be expressed as:
\begin{equation}
\label{dv_power_linear}
{Y_1}=({R}/{T_p})\int^{+\ {T_p}/{2}}_{-\ {T_p}/{2}}{{\left|r\left(t^{'}\right)\right|}^2{dt}^{'}}
\end{equation}
\begin{figure}[ht]
\centering
\includegraphics[scale=.7]{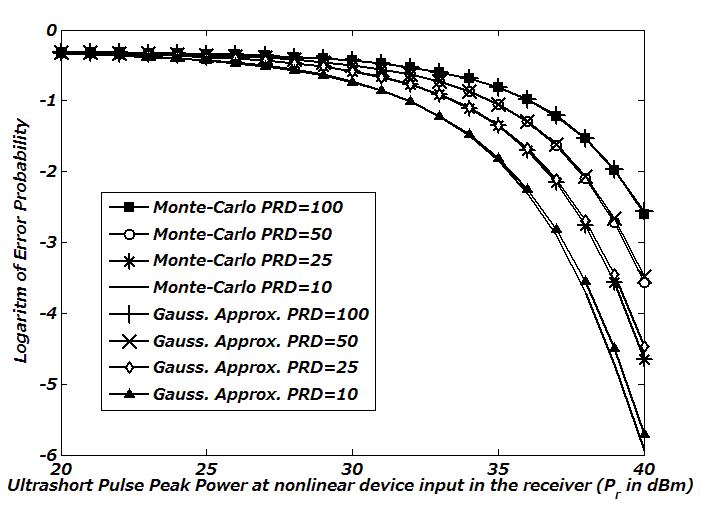}
\caption{Bit error rate of power-linear receiver for Monte-Carlo method and Gaussian approximation.\label{fig_power_lin_m_c_gauss_comp}}
\end{figure}
The result has been plotted in \figurename{ \ref{fig_power_lin_m_c_gauss_comp}}, depicting the usefulness of Gaussian approximation. Indeed, the Gaussian approximation is completely accurate and matches to the Monte-Carlo method as depicted in \figurename{ \ref{fig_power_lin_m_c_gauss_comp}}, despite the fact that the decision variable $Y_1$ in the power-linear receiver is not Gaussian random variable. Especially since, the decision variable $Y_1$ at the photodetector output could not take negative values, Gaussian distribution is not strictly approved. However, in the power-linear receiver, the Gaussian approximation for the pdf of the decision variable $Y_1$ still yields satisfactory results as it is obvious in \figurename{ \ref{fig_power_lin_m_c_gauss_comp}}.
 \\
 To verify the precision of Gaussian pdf in our problem, Monte-Carlo simulation is employed in this section utilizing the mathematical model discussed in the previous section. 
\begin{figure}[ht]
\centering
\includegraphics[scale=.8]{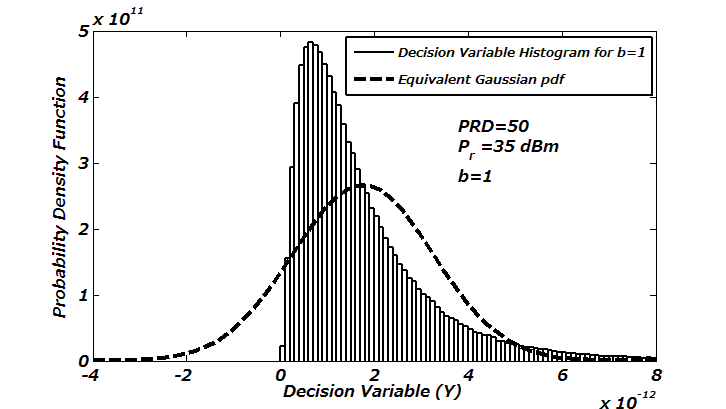}
\caption{Decision variable $Y$ histogram for bit One and comparison with Gaussian distribution.\label{fig_b_1_hist_comp_Gauss}}
\end{figure}
\begin{figure}[ht]
\centering
\includegraphics[scale=.8]{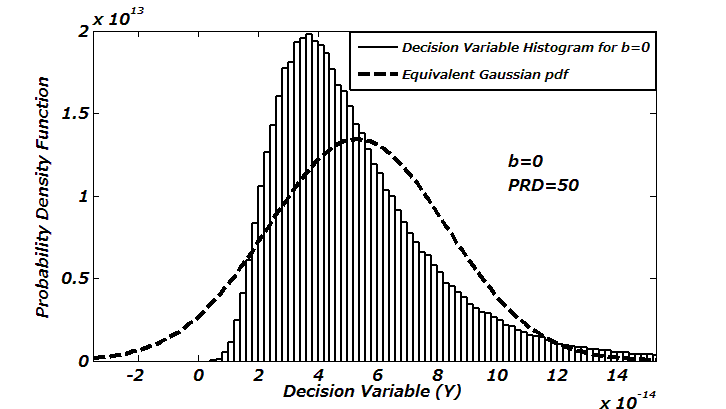}
\caption{Decision variable (Y) histogram for bit zero and comparison with Gaussian distribution.\label{fig_b_0_hist_comp_Gauss}}
\end{figure}
In Figs. \ref{fig_b_1_hist_comp_Gauss} and \ref{fig_b_0_hist_comp_Gauss} the histograms of the decision variable $Y$ are depicted for the two cases of $b=1$ and $b=0$ respectively, and its equivalent Gaussian pdf - with the same mean and variance- is also plotted for $P_r=35 \ dBm$ and $PRD=50$. It could be observed from Figs. \ref{fig_b_1_hist_comp_Gauss} and \ref{fig_b_0_hist_comp_Gauss} that Gaussian approximation is not well-fitted for our decision variable $Y$, mostly possible since $Y$ is the output of a highly nonlinear process, i.e., power-cubic process. To highlight the extent of deviation of this approximation (i.e., Gaussian pdf approximation) from it's true value, a comparison is made with the Monte-Carlo method in the case of error probability in \figurename{ \ref{fig_PE_m-c_Gauss_comp}} for two different values of $PRD$s.
\figurename{ \ref{fig_PE_m-c_Gauss_comp}} shows that Gaussian approximation leads to erroneous results especially for high peak power pulses. Among various other well-known two parameter probability distributions, such as Gamma, inverse Gaussian, Log-normal, Weibull, Erlang, Nakagami and many others, no probability distribution could be fitted to our decision variable $Y$. However, among the more advanced three parameters distributions, such as Generalized extreme value, Generalized Gamma, Generalized logistic and, etc., the Log-Pearson type III (LP3) distribution is the pdf completely fitted to the decision variable $Y$. Fitting tests are generally used to check how well the distribution fits the data samples showing the compatibility of random samples with a theoretical pdf. \tablename{ \ref{table_goodness_of_fit}} shows the result of goodness of fitting procedures with $2.5\times{10^5}$ independent samples of the decision variable $Y$ employing EasyFit software (with $P_r=35 \ dBm$, $PRD=50$ and $b=1$) for several well-known distributions  based on commonly used tests namely Kolmogorov Smirnov, Anderson Darling and Chi-Squared \cite{rayner2009smooth}.
In this table, each pdf is assigned by a rank number for a specified fitting test method that shows how well our decision variable histogram is fitted to this pdf applying the corresponding test method. Indeed, considering a specified test method, the better the fitness of a particular pdf to our decision variable histogram, the lower the rank number assigned to that pdf in the table. From \tablename{\ref{table_goodness_of_fit}}, it is clear that LP3 distribution virtually fits to our decision variable. Several typical fitting procedures using various other system parameters were performed, and nearly all had the same results and reconfirmed the goodness of LP3 distribution for the decision variable $Y$. However, only one simulation result is presented in Figs. \ref{fig_b_1_hist_comp_LP3} and \ref{fig_b_0_hist_comp_LP3} as an example.
It must be noted that the Monte-Carlo method is a powerful approach in determining the system bit error rate. However, this method is very time-consuming and requires very large computer resources. Also this approach is not feasible for error probabilities lower than almost ${10}^{-6}$ due to the long time required for simulation. In the case of our problem the Monte-Carlo method consumes much more time for higher values of $PRD$. Therefore, the LP3 model is a useful method to obtain the bit error rate of the system with a simple computer code and achieve the result in a few seconds.
\begin{table}[H]
\caption{Result of goodness of fitting tests for various well-known distributions} 
\centering
\vspace{3pt} \noindent
\begin{tabular}{|p{100pt}|p{40pt}|p{20pt}|p{40pt}|p{20pt}|p{40pt}|p{20pt}|}
\hline
\parbox{73pt}{\centering {
{\textbf{Distribution} } } } 
& \multicolumn{2}{|c|}{\parbox{37pt}{\centering 
{\textbf{Kolmogorov
\\
Smirnov}}}}
& \multicolumn{2}{|c|}{\parbox{37pt}{\centering 
{\textbf{Anderson
\\
Darling}}}} 
& \multicolumn{2}{|c|}{\parbox{43pt}{\centering 
{\textbf{Chi-Squared}}}}
\\
\cline{2-7} 
& \parbox{24pt}{\centering 
Statistic
} & \parbox{12pt}{\centering 
Rank
} & \parbox{24pt}{\centering 
Statistic
} & \parbox{12pt}{\centering 
Rank
} & \parbox{30pt}{\centering 
Statistic
} & \parbox{12pt}{\centering 
Rank
} \\
\hline
\parbox{73pt}{Log-Pearson 3}
& \parbox{24pt}{\centering 
0.00239}
& \parbox{12pt}{\centering 1}
& \parbox{24pt}{\centering 0.68473 }
& \parbox{12pt}{\centering 1 }
& \parbox{30pt}{\centering 10.837 }
& \parbox{12pt}{\centering 1} \\
\hline
\parbox{73pt}{Pearson 6}
& \parbox{24pt}{\centering 0.00388}
& \parbox{12pt}{\centering 2}
& \parbox{24pt}{\centering 2.8172}
& \parbox{12pt}{\centering 2} 
& \parbox{30pt}{\centering 23.55}
& \parbox{12pt}{\centering 2} \\
\hline
\parbox{100pt}{Inv. Gaussian(3P)}
& \parbox{24pt}{\centering 0.00391} 
& \parbox{12pt}{\centering 3} 
& \parbox{24pt}{\centering 3.3024} 
& \parbox{12pt}{\centering 3} 
& \parbox{30pt}{\centering 36.406} 
& \parbox{12pt}{\centering 3} \\
\hline
\parbox{73pt}{Lognormal(3P)}
& \parbox{24pt}{\centering 0.00725} 
& \parbox{12pt}{\centering 4} 
& \parbox{24pt}{\centering 11.234} 
& \parbox{12pt}{\centering 4} 
& \parbox{30pt}{\centering 95.342} 
& \parbox{12pt}{\centering 4} \\
\hline
\parbox{100pt}{Gen. Extreme Value}
& \parbox{24pt}{\centering 0.01357 } 
& \parbox{12pt}{\centering 5} 
& \parbox{24pt}{\centering 50.34} 
& \parbox{12pt}{\centering 5} 
& \parbox{30pt}{\centering 376.04} 
& \parbox{12pt}{\centering 5} \\
\hline
\parbox{73pt}{Dagum(4P)}
& \parbox{24pt}{\centering 0.01673} 
& \parbox{12pt}{\centering 6} 
& \parbox{24pt}{\centering 72.751} 
& \parbox{12pt}{\centering 6} 
& \parbox{30pt}{\centering 710.65} 
& \parbox{12pt}{\centering 7} \\
\hline
\parbox{73pt}{Dagum}
& \parbox{24pt}{\centering 0.01766} 
& \parbox{12pt}{\centering 7} 
& \parbox{24pt}{\centering 84.154} 
& \parbox{12pt}{\centering 8} 
& \parbox{30pt}{\centering 805.63} 
& \parbox{12pt}{\centering 8} \\
\hline
\parbox{100pt}{Log-Logistic(3P)}
& \parbox{24pt}{\centering 0.01826}
& \parbox{12pt}{\centering 8} 
& \parbox{24pt}{\centering 113.35} 
& \parbox{12pt}{\centering 10} 
& \parbox{30pt}{\centering 1177.8} 
& \parbox{12pt}{\centering 11} \\
\hline
\parbox{73pt}{Gamma(3P)}
& \parbox{24pt}{\centering 0.01932} 
& \parbox{12pt}{\centering 9} 
& \parbox{24pt}{\centering 73.032} 
& \parbox{12pt}{\centering 7} 
& \parbox{30pt}{\centering 501.36} 
& \parbox{12pt}{\centering 6} \\
\hline
\parbox{73pt}{Lognormal}
& \parbox{24pt}{\centering 0.02139} 
& \parbox{12pt}{\centering 10} 
& \parbox{24pt}{\centering 109.36} 
& \parbox{12pt}{\centering 9} 
& \parbox{30pt}{\centering 828.93} 
& \parbox{12pt}{\centering 9} \\
\hline
\parbox{73pt}{Gen. Gamma}
& \parbox{24pt}{\centering 0.02611} 
& \parbox{12pt}{\centering 11} 
& \parbox{24pt}{\centering 140.52} 
& \parbox{12pt}{\centering 11} 
& \parbox{30pt}{\centering 925.61} 
& \parbox{12pt}{\centering 10} \\
\hline
\parbox{73pt}{Gamma}
& \parbox{24pt}{\centering 0.02757} 
& \parbox{12pt}{\centering 12} 
& \parbox{24pt}{\centering 211.63} 
& \parbox{12pt}{\centering 12} 
& \parbox{30pt}{\centering 1551.6} 
& \parbox{12pt}{\centering 12} \\
\hline
\parbox{73pt}{Gen. Logistic}
& \parbox{24pt}{\centering 0.02828} 
& \parbox{12pt}{\centering 13} 
& \parbox{24pt}{\centering 211.88} 
& \parbox{12pt}{\centering 13}
& \parbox{30pt}{\centering 1854.0} 
& \parbox{12pt}{\centering 14} \\
\hline
\parbox{73pt}{Inv. Gaussian}
& \parbox{24pt}{\centering 0.03005} 
& \parbox{12pt}{\centering 14} 
& \parbox{24pt}{\centering 323.7} 
& \parbox{12pt}{\centering 15} 
& \parbox{30pt}{\centering 2115.8} 
& \parbox{12pt}{\centering 15} \\
\hline
\parbox{73pt}{Log-Logistic}
& \parbox{24pt}{\centering 0.03326} 
& \parbox{12pt}{\centering 15} 
& \parbox{24pt}{\centering 216.26} 
& \parbox{12pt}{\centering 14} 
& \parbox{30pt}{\centering 1664.8} 
& \parbox{12pt}{\centering 13} \\
\hline
\parbox{73pt}{Weibull~(3P)}
& \parbox{24pt}{\centering 0.03861} 
& \parbox{12pt}{\centering 16} 
& \parbox{24pt}{\centering 440.68}
& \parbox{12pt}{\centering 16} 
& \parbox{30pt}{\centering 3355.0}
& \parbox{12pt}{\centering 16} \\
\hline
\parbox{73pt}{Weibull}
& \parbox{24pt}{\centering 0.05004} 
& \parbox{12pt}{\centering 17} 
& \parbox{24pt}{\centering 793.76} 
& \parbox{12pt}{\centering 17} 
& \parbox{30pt}{\centering 4883.3} 
& \parbox{12pt}{\centering 17} \\
\hline
\parbox{73pt}{Pearson~5}
& \parbox{24pt}{\centering 0.07089} 
& \parbox{12pt}{\centering 18} 
& \parbox{24pt}{\centering 1151.2} 
& \parbox{12pt}{\centering 18} 
& \parbox{30pt}{\centering 8080.8} 
& \parbox{12pt}{\centering 18} \\
\hline
\parbox{73pt}{Pearson~5~(3P)}
& \parbox{24pt}{\centering 0.08537} 
& \parbox{12pt}{\centering 10} 
& \parbox{24pt}{\centering 1703.4} 
& \parbox{12pt}{\centering 19} 
& \parbox{30pt}{\centering 11043.0} 
& \parbox{12pt}{\centering 21} \\
\hline
\parbox{73pt}{Normal}
& \parbox{24pt}{\centering 0.10397} 
& \parbox{12pt}{\centering 20} 
& \parbox{24pt}{\centering 2750.2} 
& \parbox{12pt}{\centering 21} 
& \parbox{30pt}{\centering 21781.0} 
& \parbox{12pt}{\centering 23} \\
\hline
\parbox{73pt}{Logistic}
& \parbox{24pt}{\centering 0.10711} 
& \parbox{12pt}{\centering 21}
& \parbox{24pt}{\centering 2442.1} 
& \parbox{12pt}{\centering 20} 
& \parbox{30pt}{\centering 19919.0} 
& \parbox{12pt}{\centering 22} \\
\hline
\parbox{73pt}{Erlang}
& \parbox{24pt}{\centering 0.1109} 
& \parbox{12pt}{\centering 22} 
& \parbox{24pt}{\centering 3592.6} 
& \parbox{12pt}{\centering 22} 
& \parbox{30pt}{\centering 8292.7} 
& \parbox{12pt}{\centering 19} \\
\hline
\parbox{73pt}{Erlang~(3P)}
& \parbox{24pt}{\centering 0.12267} 
& \parbox{12pt}{\centering 23} 
& \parbox{24pt}{\centering 4806.9} 
& \parbox{12pt}{\centering 23} 
& \parbox{30pt}{\centering 10322.0} 
& \parbox{12pt}{\centering 20} \\
\hline

\end{tabular}
\vspace{2pt}
\label{table_goodness_of_fit}
\end{table}
\section{Statistical Modeling of the Decision Variable ($Y$) Based on LP3 Distribution \label{sec_ver_Dv_LP3}}
As discussed in the section \ref{sec_ver_DV}, the general form of the decision variable $Y$ is well modeled with LP3 distribution, however in finding the exact mathematical form of the required LP3 pdf we need to obtain three unknown characteristic parameters of the LP3 pdf. In general LP3 pdf is expressed as follows \cite{singh1998entropy,koutrouvelis2000comparison}:
\begin{equation}
\label{LP3_pdf}
f^{(b)}_Y\left(y\right)=\frac{1}{y\left|{{\beta}_b}\right|\Gamma\left({\alpha}_b\right)}\times {\left[\frac{Ln\left(y\right)-{\gamma}_b}{{\beta}_b}\right]}^{\left({\alpha}_b-1\right)}\times e^{-\ \left(\frac{Ln\left(y\right)-{\gamma}_b}{{\beta}_b}\right)}\ \ \ \ \ ;b=0,1 
\end{equation}
Also, the cumulative distribution function (cdf) of LP3 random variable can be expressed as:
\begin{align}
\label{LP3_cdf}
F^{(b)}_Y\left(y\right)=\{ \begin{array}{c}
{P\left({\alpha}_b\ ,\ \ \frac{Ln\left(y\right)-\ {\gamma}_b}{{\beta}_b}\ \right)\ \ \ if\ {\beta}_b>0} \\
{Q\left({\alpha}_b\ ,\ \ \frac{Ln\left(y\right)-\ {\gamma}_b}{{\beta}_b}\ \right)\ \ \ if\ {\beta}_b<0}\end{array}
\end{align}
Where ${\alpha}_b>0$, ${\beta}_b\ne0$ and ${\gamma}_b$ are unknown parameters which must be characterized to completely express the expressions for LP3 pdf and cdf. ${\Gamma}(\centerdot)$ and $Ln(\centerdot)$ denote Gamma function and natural logarithm function, respectively. Also $P$ is the lower incomplete gamma function defined as $P\left(a,x\right)=({1}/{{\Gamma}\left(a\right)})\int^x_0{e^{-t}t^{a-1}dt}$ and $Q$ is the upper incomplete gamma function defined as $Q\left(a,x\right)=1-P\left(a,x\right)=({1}/{{\Gamma}\left(a\right)})\int^{\infty }_x{e^{-t}t^{a-1}dt}$. Furthermore, the moments of a random variable $Y$ with LP3 distribution can be expressed as \cite{koutrouvelis2000comparison}:
\begin{equation}
\label{moments_of_LP3}
E\left\{{Y^n}/{\left(transmitted\ bit=b\right)}\right\}\triangleq {{{}\mu}^{(b)}_n}\}=e^{n{\gamma}_b}\times {\left(1-n{\beta}_b\right)}^{-{\alpha}_b} \ ; \ b=0,1
\end{equation}
Based on the system model presented in section \ref{sec_system_math_model}, the first three moments, ${\mu}^{(b)}_1$, ${\mu}^{(b)}_2$ and ${\mu}^{(b)}_3$ of the decision variable $Y$ are obtained analytically in appendices \ref{app_mom1}, \ref{app_mom2} and \ref{app_mom3} respectively. The first order moment can be expressed as:
\begin{equation}
\label{mom1_b_matn}
{\mu{}}_1^{(b)}=\frac{Rk{\Gamma{}}^2}{PRD}(48{\sigma{}}_0^6 PRD +72{\sigma{}}_0^4 b  P_r+12{\sigma{}}_0^2 b P_r^2+0.55 b{P_r}^3) \ \ ; b=0,1
\end{equation}
The second order moment can be expressed as:
\begin{equation}
\label{mom2_b_matn}
{\mu{}}_2^{(b)}=\
\frac{R^2k^2{\Gamma{}}^4}{PRD^2}\times ({{\mu{}}_{2,1}^{(b)}+{\mu{}}_{2,2}^{(b)}+{\mu{}}_{2,3}^{(b)}+{\mu{}}_{2,4}^{(b)}}) \ \ ; b=0,1
\end{equation}
Where:
\begin{equation}
\label{mom2_b_matn_1}
{\mu{}}_{2,1}^{(b)}=2304{\sigma{}}_0^{12}\times{}PRD^2+35834{\sigma{}}_0^{12}\times{}PRD
\end{equation}
\begin{equation}
\label{mom2_b_matn_2}
{\mu{}}_{2,2}^{(b)}=6912{\sigma{}}_0^{10} b P_r\times{}PRD+1152{\sigma{}}_0^8 b P_r^2\times{}PRD
\end{equation}
\begin{equation}
\label{mom2_b_matn_3}
{\mu{}}_{2,3}^{(b)}=52.8{\sigma{}}_0^6 b P_r^3\times{}PRD+106320{\sigma{}}_0^{10} b P_r \times{}PRD
\end{equation}
\begin{equation}
\label{mom2_b_matn_4}
{\mu{}}_{2,4}^{(b)}=47319{\sigma{}}_0^8 b P_r^2
+8661.7{\sigma{}}_0^6 b P_r^{3\
}+691.2{\sigma{}}_0^4 b P_r^4+24.15{\sigma{}}_0^2 b P_r^5
 +0.3025 b P_r^6
\end{equation}
The third order moment can be expressed as:
\begin{equation}
\label{mom3_b_matn}
{\mu{}}_3^{(b)}={{\mu{}}_{3,1}^{(b)}+{\mu{}}_{3,2}^{(b)}+{\mu{}}_{3,3}^{(b)}+{\mu{}}_{3,4}^{(b)}} \ \ ; b=0,1
\end{equation}
In which:
\begin{equation}
\label{mom3_b_matn_1}
{\mu{}}_{3,1}^{(b)}=R^3k^3{\Gamma{}}^6\times{}\left(110592{\sigma{}}_0^{18}\right)
\end{equation}
\begin{equation}
\label{mom3_b_matn_2}
{\mu{}}_{3,2}^{(b)}=\frac{R^3k^3{\Gamma{}}^6}{PRD}\times{}\left(\begin{array}{l}5.16\times{}{10}^6{\sigma{}}_0^{18}+4.977\times{}{10}^5 b P_r{\sigma{}}_0^{16}
\\
+8.3\times{}{10}^4 b P_r^2{\sigma{}}_0^{14}+3.8\times{}{10}^3 b P_r^3{\sigma{}}_0^{12}\end{array}\right)
\end{equation}
\begin{equation}
\label{mom3_b_matn_3}
{\mu{}}_{3,3}^{(b)}=\frac{R^3k^3{\Gamma{}}^6}{PRD^2}\times{}\left(\begin{array}{l}1.0538\times{}{10}^8{\sigma{}}_0^{18}+2.308\times{}{10}^7 b P_r{\sigma{}}_0^{16}+
\\
8.133\times{}{10}^6 b P_r^2{\sigma{}}_0^{14}
+1.306\times{}{10}^6 b P_r^3{\sigma{}}_0^{12}+ \\
9.956\times{}{10}^4 b P_r^4{\sigma{}}_0^{10}+3.479\times{}{10}^3 b P_r^5{\sigma{}}_0^8
\\
+43.56 b P_r^6{\sigma{}}_0^6\end{array}\right)
\end{equation}
\begin{equation}
\label{mom3_b_matn_4}
{\mu{}}_{3,4}^{(b)}=\frac{R^3k^3{\Gamma{}}^6}{PRD^3}\times{}\left(\begin{array}{l}4.671\times{}{10}^8 b P_r{\sigma{}}_0^{16}+3.241\times{}{10}^8 b P_r^2{\sigma{}}_0^{14}+ \\
1.027\times{}{10}^8 b P_r^3{\sigma{}}_0^{12}
+1.647\times{}{10}^7 b P_r^4{\sigma{}}_0^{10}+ \\
1.451\times{}{10}^6 b P_r^5{\sigma{}}_0^8+7.232\times{}{10}^4 b P_r^6{\sigma{}}_0^6
\\
+2.014\times{}{10}^3 b P_r^7{\sigma{}}_0^4+28.97 b P_r^8{\sigma{}}_0^2+0.1664 b P_r^9\end{array}\right)
\end{equation}
Using equation \eqref{moments_of_LP3}, it is shown in appendix \ref{app_mom_to_param} if the first 3 moments of LP3 random variable are known, one can easily find ${\alpha}_b$, ${\beta}_b$ and ${\gamma}_b$. If we denote ${\mu}^{(b)}_1$, ${\mu}^{(b)}_2$ and ${\mu}^{(b)}_3$ as first, second and third  moments of $Y$, then the below expressions are valid as explained in appendix \ref{app_mom_to_param}:
\begin{equation}
\label{LP3_mom_to_param1}
{\frac{3Ln\left(1-{\beta}_b\right)-Ln\left(1-3{\beta}_b\right)}{2Ln\left(1-{\beta}_b\right)-Ln(1-2{\beta}_b)}}={\frac{Ln\left({\mu}^{(b)}_3\right)-3Ln\left({\mu}^{(b)}_1\right)}{Ln\left({\mu}^{(b)}_2\right)-2Ln \left({\mu}^{(b)}_1\right)}} 
\end{equation}
\begin{equation}
\label{LP3_mom_to_param2}
{\alpha}_b=\frac{Ln\left({\mu}^{(b)}_2\right)-2Ln\left({\mu}^{(b)}_1\right)}{2Ln\left(1-{\beta}_{b}\right)-Ln(1-2{\beta}_{b})}
\end{equation}
\begin{equation}
\label{LP3_mom_to_param3}
{\gamma}_b=Ln\left({\mu}^{(b)}_1\right)+{\alpha}_b\times Ln(1-{\beta}_b)
\end{equation}
Indeed, the nonlinear equation \eqref{LP3_mom_to_param1} first has to be solved in order to obtain ${\beta}_b$ parameter. Knowing ${\beta}_b$, ${\alpha}_b$ is obtained readily in equation \eqref{LP3_mom_to_param2}, and finally ${\gamma}_b$ is obtained from equation \eqref{LP3_mom_to_param3}.
\section{Effects of Shot and Thermal Noises \label{sec_sh_th_effect}}
In this section, effects of shot noise of the receiver's photodetector and thermal noise of the front end electronic stage of the receiver are taken into account and the statistics of decision variable is derived. We denote $Y_{sh,th}$ to be the decision variable that considers the effects of shot and thermal noises and can be defined at the output of the receiver's sampler as follows:
\begin{equation}
\label{shot_noise_general}
Y_{sh,th}={\left.i_{shot-noise}\left(t\right)+i_{thermal}\left(t\right)\right|}_{@t=(nT_b+\frac{T_p}{2})}=i_{shot}+i_{thermal}
\end{equation}
In the above equation, $i_{shot-noise}(t)$ is the shot noise current at the output of the photodetector and  $i_{thermal}\left(t\right)$ is the current added to the shot noise due to thermal noise of the electronic front end of the receiver. Indeed, $i_{shot-noise}(t)$ includes the statistical properties of the added Gaussian noise to the signal due to ASE and shot noise of the photodetector. For obtaining the statistics of the final decision variable $Y_{sh,th}$, we first consider its statistics conditioned on ASE noise. Since the random variable $Y_{sh,th}$ includes $Y$, then the condition has to be applied on $Y$. The cdf of random variable $Y_{sh,th}$ can be written as:
\begin{equation}
\label{cdf_sh_th_barhasbe_pdf_no_sh_th}
F^{(b)}_{Y_{sh,th}}\left(x\right)\triangleq P\left(Y_{sh,th}\le x\right)=\int^{+\infty }_0{P\left(Y_{sh,th}\le x\ |\ Y=y\right)}f^{(b)}_Y\left(y\right)dy\ \ ;\ \  b=0,1 
\end{equation}
Conditioned on ($Y=y$), $i_{shot}$ and $i_{thermal}$ are two independent Gaussian random variables with $y$ and zero mean and variances $\sigma_{shot}^2={{(2q_ey)}/{T_p}}$ and $\sigma_{thermal}^2=(4K_BT_r)/(R_L T_p)$, respectively, where $K_B$ is the Boltzman constant and $T_r$ is the absolute temperature of the receiver in Kelvin and $R_L$ is the equivalent load resistance in the front end stage of the electronic part of the receiver \cite{govind2010fiber}. The bandwidth of the receiver's photodetector and electronic circuit are assumed to be identical and equal to $BW=1/T_p$ . The exact statistics of  $i_{shot}$ conditioned on ($Y=y$) follows a Poisson distribution \cite{govind2010fiber}, but for the sake of mathematical simplicity we consider it to be approximately Gaussian without any loss of accuracy \cite{govind2010fiber}. Accordingly, because of statistical independence between the shot and thermal noises, the random variable $Y_{sh,th}$, conditioned on ($Y=y$), is a normal random variable with a mean value equal to $y$ and variance equal to $((2q_ey)/T_p+(4K_BT_r)/(R_L T_p))$. Therefore, the cdf of random variable $Y_{sh,th}$, can be expressed as:
\begin{equation}
\label{Y_sh_th_gaussian_pdf_cond_on_y}
\left({Y_{sh,th}}/{Y=y}\right)\sim Normal\left(y,\left(\frac{2q_e}{T_p}y+\frac{4K_BT_r}{R_L T_p}\right)\right)
\end{equation}
\begin{equation}
\label{Y_sh_th_gaussian_cdf_cond_on_y}
P\left({Y_{sh,th}\le x}/{Y=y}\right)\triangleq u(y)=\Phi \left((x-y)\times {\left(\frac{2q_e}{T_p}y+\frac{4K_BT_r}{R_L T_p}\right)}^{-\frac{1}{2}}\right) 
\end{equation}
Using equations \eqref{cdf_sh_th_barhasbe_pdf_no_sh_th} and \eqref{Y_sh_th_gaussian_cdf_cond_on_y}, the cdf of $Y_{sh,th}$ can be written as:
\begin{equation}
\label{cdf_barhasbe_u}
F^{(b)}_{Y_{sh,th}}\left(x\right)=\int_{0}^{+\infty{}}{u(y)f^{(b)}_Y\left(y\right)dy}\  ;\  b=0,1
\end{equation}
Where $\Phi(.)$ is the cumulative distribution function (cdf) of the standard normal distribution  \cite{simon2007probability} and is defined as $\Phi{}\left(x\right)\triangleq{}\frac{1}{\sqrt{2\pi{}}}\int_{-\infty{}}^{x}e^{-{t^2}/{2}}
dt=\frac{1}{2}\left[1+erf{\left(\frac{x}{\sqrt{2}}\right)}\right]$. For the derivation of the error probability, including the effect of shot and thermal noises, the integral in equation \eqref{cdf_barhasbe_u} must be calculated numerically. Whereas, if we evoke $F_Y^{(b)}(.)$ which is numerically simpler than $f_Y^{(b)}(.)$, and using the mathematical identity ($\int_{}^{}{udv}=uv-\int_{}^{}{vdu}$) supposing $dv=f_Y^{(b)}(y)dy$ ($v=F_Y^{(b)}(y)$) and also noting that $u(+\infty)=\Phi(-\infty)=0$ and $v(0)=F_Y^{(b)}(0)=0$ and obtaining  ${du}/{dy}={u^{'}}(y)$ as:
\begin{equation}
\label{u_prim}
{u^{'}}\left(y\right)=\frac{-1}{\sqrt{2\pi}}{\left[\frac{q_e}{T_p}\left(x+y\right)+\frac{4K_BT_r}{R_L T_p}\right] {\left(\frac{2q_e}{T_p}y+\frac{4K_BT_r}{R_L T_p}\right)}^{-{3}/{2}}}e^{-W(x,y)}
\end{equation}
Where $W(x,y)$ in \eqref{u_prim} is:
\begin{equation}
\label{W_x_y}
W(x,y)=\frac{{\left(x-y\right)}^2}{2\left(\frac{2q_e}{T_p}y+\frac{4K_BT_r}{R_L T_p}\right)}
\end{equation}
Then, substituting ($du={u^{'}}(y)dy$) in ($\int_{}^{}{vdu}$) from \eqref{u_prim}, equation \eqref{cdf_barhasbe_u} can be shown to be equal to:
\begin{equation}
\label{cdf_barhasbe_u_prim}
F^{(b)}_{Y_{sh,th}}\left(x\right)=\int^{+\infty }_0{(-{u^{'}}(y))F^{(b)}_Y\left(y\right)dy} \ \ ;\ b=0,1
\end{equation}
Substituting equation \eqref{LP3_cdf} into equation \eqref{cdf_barhasbe_u_prim} and employing numerical integration, $F_{Y_{sh,th}}^{(b)}(x)$ can be readily obtained.
\section{Error Probability and Optimum Threshold \label{sec_error_prob_and_opt_det}}
The error probability of the system can be expressed as follows:
\begin{equation}
\label{error_prob_general}
PE=\frac{1}{2}PE\left({Detecting \ One}/{Sending \ Zero}\right)\
+\frac{1}{2}PE\left({Detecting \ Zero}/{Sending \ One}\right)
\end{equation}
Note that the cdf of the decision variable for two cases of with and without shot and thermal noise effects were calculated in previous sections. And if the threshold level (i.e., $Th$) is known and fixed, the error probability in equation \eqref{error_prob_general} can be written as:
\begin{equation}
\label{PE_barhasbe_cdfha}
PE=\frac{1}{2}\left(1-F_{\xi{}}^{(0)}\left(Th\right)\right)+\frac{1}{2}\left(F_{\xi{}}^{(1)}\left(Th\right)\right)\
\ ;\ \ \xi{}=Y,Y_{sh,th}
\end{equation}
Where in equation \eqref{PE_barhasbe_cdfha}, $F_{\xi}^{(0)}$ and $F_{\xi}^{(1)}$ are the cdf of the decision variable assuming bit zero and bit one being transmitted, respectively. Also, the variable  $\xi$ can take on two values namely $Y$ given in equation \eqref{LP3_cdf} and $Y_{sh,th}$ expressed in equation \eqref{cdf_barhasbe_u_prim} denoting decision variable with shot and thermal noises exclusion and inclusion, respectively. In obtaining the optimum threshold, $Th$ must be specified such that the expression for $PE$ in equation \eqref{PE_barhasbe_cdfha} is minimized { (${\left.PE\right\vert{}}_{min}={\left.PE\right\vert{}}_{Th=Th_{OPT}}$)}.  In the next section, the error probability curves are numerically calculated based on minimizing equation \eqref{PE_barhasbe_cdfha} in order to find the optimum threshold and its equivalent error probability.

\section{Numerical Results \label{sec_Num_Res}}
\begin{figure}[ht]
\centering
\includegraphics[scale=.7]{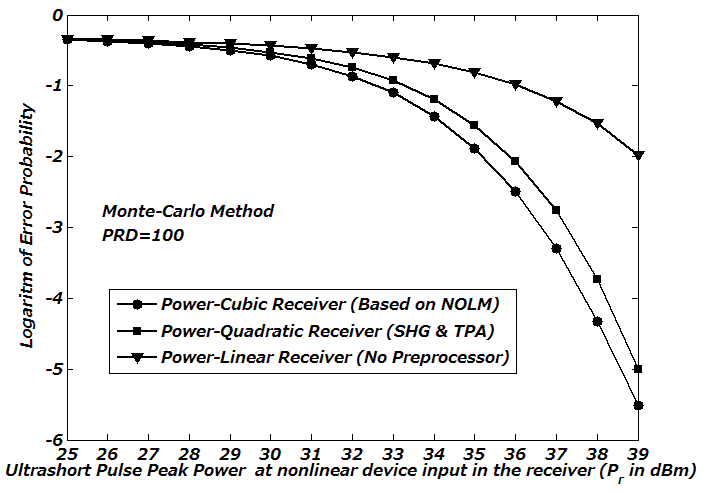}
\caption{Error probability comparison of different receiver types based on Monte-Carlo simulation.\label{fig_mc_pl_pq_pc}}
\end{figure}

The parameters used in our numerical error probability calculations are shown in \tablename{ \ref{table_numeric_values}}. First, a Monte-Carlo simulation with ${10}^7$ independent samples of decision variable $Y$ is performed to serve the performance comparison of power-cubic receiver against power-quadratic (SHG and TPA devices) and power-linear (no nonlinear preprocessor) counterparts. Recalling decision variable of the power linear receiver $Y_1$ in equation \eqref{dv_power_linear}, for a power-quadratic receiver, the decision variable $Y_2$ can be expressed as:
  \begin{equation}
 \label{dv_power_quadratic}
 {{Y_2}=({R{\vartheta}}/{T_p})\int^{+\ {T_p}/{2}}_{-\  {T_p}/{2}}{{\left|r\left(t^{'}\right)\right|}^4{dt}^{'}}}
 \end{equation}
  Where ${\vartheta}$ in \eqref{dv_power_quadratic} is a device parameter related to the physical structure of TPA or SHG. Due to the fact that we neglect shot and thermal noises in our analysis in \figurename{ \ref{fig_mc_pl_pq_pc}}, the value of ${\vartheta}$ is irrelevant in our simulations in this figure. By letting ${\vartheta}=1$ the result of the Monte-Carlo simulation is depicted in \figurename{ \ref{fig_mc_pl_pq_pc}}. \figurename{ \ref{fig_mc_pl_pq_pc}} shows performance superiority of the power-cubic receiver over the other two receivers. It is worth noting that many other simulations with various parameters were carried out and all resulted in demonstrating the performance superiority of the power-cubic receiver. However, for the sake of brevity we only presented one of the many cases in \figurename{ \ref{fig_mc_pl_pq_pc}}.

\begin{figure}[ht]
\centering
\includegraphics[scale=.7]{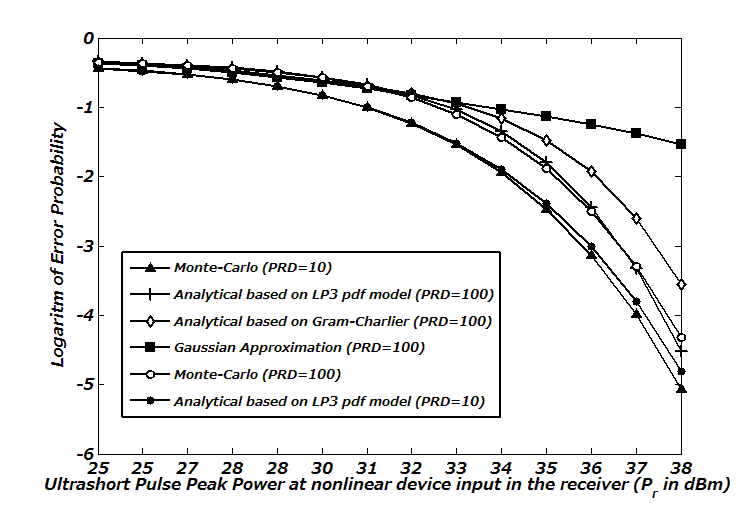}
\caption{Error probability comparison of Monte-Carlo approach and LP3 distribution model for power-cubic preprocessor.\label{fig_mc_LP3_Gauss_Gram_charl}}
\end{figure}

\figurename{ \ref{fig_mc_LP3_Gauss_Gram_charl}} shows the error probability curves based on the analytical LP3 model of the decision variable $Y$ and the Monte-Carlo approach neglecting shot and thermal noises. Also the results of Gaussian approximation and Gram-Charlier analytical method -all for a power-cubic nonlinear receiver- presented in \cite{matinfar2011performance} are depicted. It is clear that our proposed LP3 analytical model is an accurate model. In \cite{matinfar2011performance}, the Gram-Charlier method whose pdf is approximated based on the first three moments of the random variable, is employed in order to model the pdf of the power-quadratic receiver. However, when the Gram-Charlier method is used for the case of a power-cubic receiver, according to our several comparisons with various system parameters, this method is no longer accurate and the proposed LP3 model, as it is shown in \figurename{ \ref{fig_mc_LP3_Gauss_Gram_charl}}, is more favorable.
\begin{figure}[ht]
\centering
\includegraphics[scale=.7]{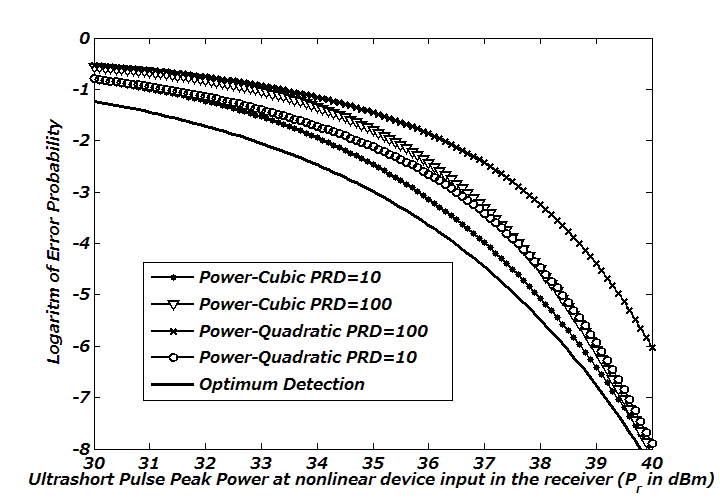}
\caption{Error Probability comparison with respect to the received power ignoring shot and thermal noises.\label{fig_no_sh_th}}
\end{figure}

\begin{figure}[ht]
\centering
\includegraphics[scale=.65]{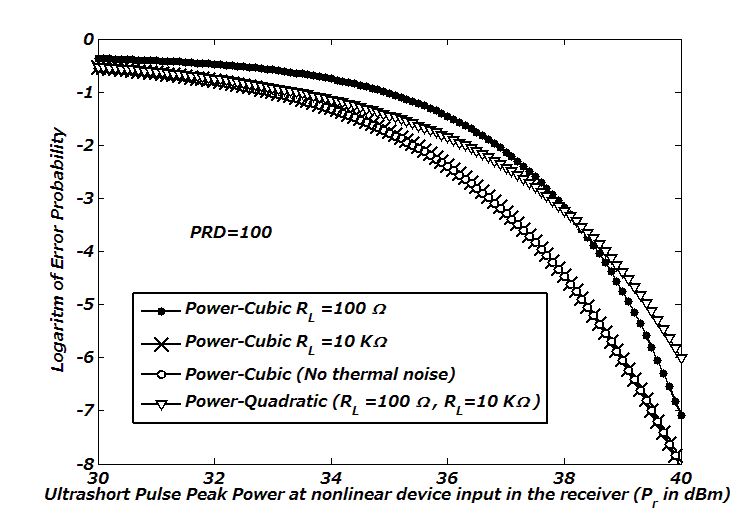}
\caption{Error Probability comparison with respect to the received power (ASE, shot and thermal noises are valid).\label{fig_sh_th_RL_mokhtalef}}
\end{figure}

\begin{figure}[ht]
\centering
\includegraphics[scale=.7]{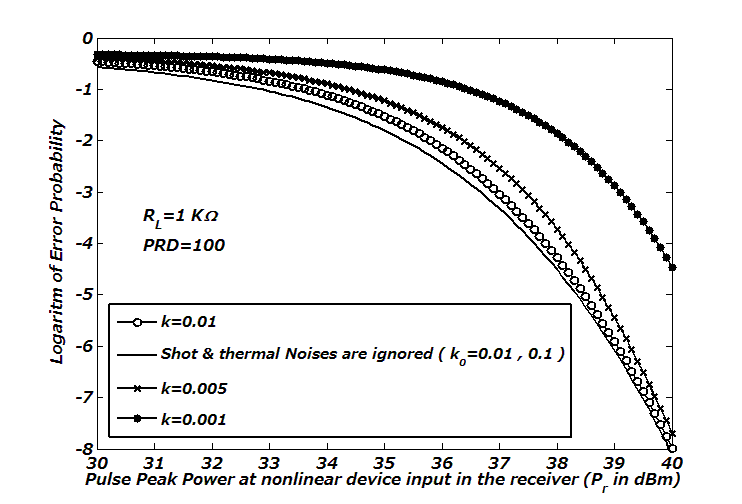}
\caption{Effect of power-cubic device parameter k in the performance of the system.\label{fig_k_mokhtalef}}
\end{figure}

\begin{figure}[ht]
\centering
\includegraphics[scale=.7]{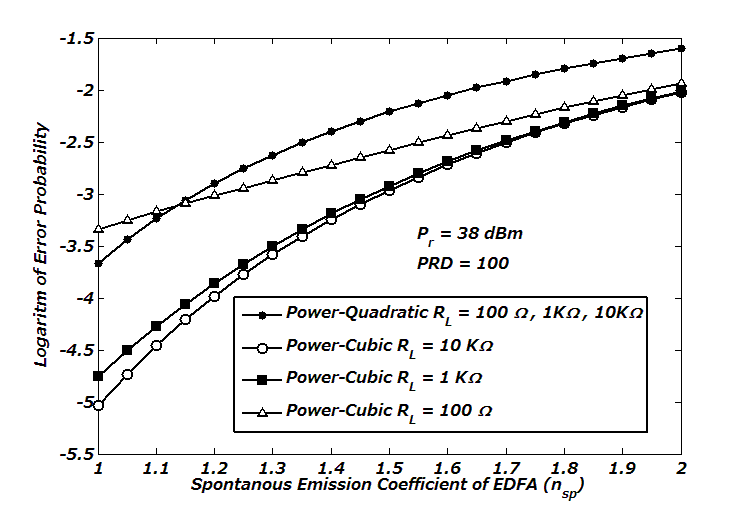}
\caption{Error probability of the system with respect to ASE noise power.\label{fig_ase_power}}
\end{figure}

In \figurename{ \ref{fig_no_sh_th}}, the error probability of an ultrafast digital lightwave communication system is depicted considering ASE noise of EDFA and ignoring shot and thermal noises. This condition is valid when ASE is the dominant noise in the system while shot and thermal noises are negligible. Also, the error probability for power-quadratic receiver (thin SHG crystal) based on Gram-Charlier approach employed in \cite{matinfar2011performance} is shown for comparison. Physical parameters of SHG used here, are also specified in \tablename{ \ref{table_numeric_values}}. The curve of error probability for ultrafast optimum detection (employing an ultrafast photodetector without any nonlinear preprocessor) is also shown in the same figure \cite{salehi1990coherent}. As expected based on our Monte-Carlo simulations, neglecting shot and thermal noises, power-cubic receiver has superior performance than power-quadratic receiver. In \figurename{ \ref{fig_sh_th_RL_mokhtalef}} the effects of shot and thermal noises are included. For a low thermal noise condition (i.e., $R_L= 10 K\Omega$) the power-cubic thresholder outperforms the power-quadratic. However, for a high thermal noise situation (i.e., $R_L=100 \Omega$) in low power regime ($P_r < 38 dBm$) the SHG device has superior performance. On the contrary, for a high power regime ($P_r > 38 dBm$) the power-quadratic device has better performance due to the dominance of the ASE noise in high power regime. Indeed, due to the NOLM design in \cite{kravtsov2007simple} with a low value of power transmittance parameter ($k=0.01$), in the low power regime, the output power of NOLM is lower than SHG and as a result, NOLM has a lower performance than SHG.    
\\
Also, in \figurename{ \ref{fig_k_mokhtalef}} the effect of the power transmittance parameter of the device ($k$) on the system performance is shown and as mathematical formulations suggest, higher values of $k$ eventuates a more robust operation of the receiver when encountering the destructive effects of shot and thermal noises. Hence, in the design of the power cubic device this point must be taken into consideration. In \figurename{ \ref{fig_ase_power}}, for a fixed received peak power at the receiver ($P_r=38 \ dBm$) the performance of the power-cubic and power-quadratic systems are compared versus ASE noise power. From this figure it can be observed that for high thermal noise conditions (i.e., $R_L=100\Omega$) and low ASE conditions, the SHG device outperforms the power-cubic device but when ASE power is increased the power-cubic based system shows a superior performance when compared to SHG. Also in lower thermal noise conditions (i.e., $R_L=10 K\Omega$) the power-cubic preprocessor has greater performance for any values of ASE power level. As a result, in the device design and implementation of the NOLM, greater value of $k$ must be taken into account.  

\section{Conclusion\label{sec_conclusion}}
To conclude, the power-cubic NOLM based nonlinear preprocessor in an ultrafast digital lightwave communication system is mathematically investigated. We began by modeling the mathematical structure of the receiver, and the corresponding error probability curves were obtained using Monte-Carlo simulations. We also show the low accuracy of Gaussian approximation using Monte-Carlo simulations. Afterward, the goodness of Log-Pearson type III (LP3) distribution for characterization of the decision variable is shown based on matching the decision variable histograms to various known distributions. In specifying the parameters of LP3 distribution, the first three moments of the decision variable are obtained, and the parameters of LP3 were derived based on those three moments. We followed our analysis by adding the effects of shot and thermal noises to the receiver and an expression for the distribution of decision variable is also obtained. Finally based on our mathematical and statistical evaluations, performance and error probability of the ultrafast digital lightwave system were numerically calculated and error probability curves for various conditions are compared. Also the performance of power-cubic nonlinear preprocessor is compared with a power-quadratic preprocessor (SHG crystal) in various conditions and the performance superiority of power-cubic in various scenarios, i.e., low thermal noise conditions, high received power conditions and high ASE noise power conditions, is demonstrated.


%

\appendices
\section{Derivation of LP3 Parameters Based on Moments of $Y$ \label{app_mom_to_param}}
Here the method of derivation of equations \eqref{LP3_mom_to_param1}, \eqref{LP3_mom_to_param2} and \eqref{LP3_mom_to_param3} are explained in detail. Based on \eqref{moments_of_LP3}, we have:
\begin{equation}
\label{mom1A1}
{\mu{}}_1^{(b)}=e^{{\gamma{}}_b}\times{}{\left(1-{\beta{}}_b\right)}^{-{\alpha{}}_b}
\end{equation}
\begin{equation}
\label{mom2A1}
{\mu{}}_2^{(b)}=e^{2{\gamma{}}_b}\times{}{\left(1-2{\beta{}}_b\right)}^{-{\alpha{}}_b}
\end{equation}
\begin{equation}
\label{mom3A1}
{\mu{}}_3^{(b)}=e^{3{\gamma{}}_b}{\times{}\left(1-3{\beta{}}_b\right)}^{-{\alpha{}}_b}
\end{equation}
Taking natural logarithm of both sides of equations \eqref{mom1A1}, \eqref{mom2A1} and \eqref{mom3A1}:
\begin{equation}
\label{Ln_mom1_A1}
Ln({\mu{}}_1^{(b)})={\gamma{}}_b-{\alpha{}}_bLn\left(1-{\beta{}}_b\right)
\end{equation}
\begin{equation}
\label{Ln_mom2_A1}
Ln({\mu{}}_2^{(b)})=2{\gamma{}}_b-{\alpha{}}_bLn\left(1-2{\beta{}}_b\right)
\end{equation}
\begin{equation}
\label{Ln_mom3_A1}
Ln({\mu{}}_3^{(b)})=3{\gamma{}}_b-{\alpha{}}_bLn\left(1-3{\beta{}}_b\right)
\end{equation}
$\gamma_b$ is eliminated by simultaneously solving \eqref{Ln_mom1_A1} and \eqref{Ln_mom2_A1}. As a result $\alpha_b$ is expressed as follows:
\begin{equation}
\label{alpha_mom1-mom2_beta_A1}
{\alpha{}}_b=\frac{Ln\left({\mu{}}_2^{(b)}\right)-2Ln\left({\mu{}}_1^{(b)}\right)}{2Ln\left(1-{\beta{}}_b\right)-Ln\left(1-2{\beta{}}_b\right)}
\end{equation}
Furthermore, $\alpha_b$ can be written differently by replacing $\gamma_b$ from \eqref{Ln_mom1_A1} in \eqref{Ln_mom3_A1}:
\begin{equation}
\label{alpha_mom1-mom3_beta_A1}
{\alpha{}}_b=\frac{Ln({\mu{}}_3^{(b)})-3Ln({\mu{}}_1^{(b)})}{3Ln\left(1-{\beta{}}_b\right)-Ln(1-3{\beta{}}_b)}
\end{equation}
Equating \eqref{alpha_mom1-mom2_beta_A1} and \eqref{alpha_mom1-mom3_beta_A1}, the nonlinear \eqref{LP3_mom_to_param1} is deduced. This equation has to be numerically solved to give $\beta_b$ value. Knowing $\beta_b$, $\alpha_b$ would be readily obtained as in \eqref{alpha_mom1-mom2_beta_A1} and then, using \eqref{Ln_mom1_A1}, $\gamma_b$ is obtained as expressed in \eqref{LP3_mom_to_param3}.
\section{Deriving First Order Moment \label{app_mom1}}
In this appendix, the first order moment of the random variable expressed in \eqref{DV main equation} is obtained. Assuming $b=1$, the expectation value of expression \eqref{DV main equation} is obtained as follows:
\begin{align}
\label{mom1_ebarate1_B}
&{{\mu{}}_1^{(1)}\triangleq{}E\left\{{Y}/{b=1}\right\}=\frac{Rk{\Gamma{}}^2}{T_p}\int_{-\
\frac{T_p}{2}}^{+\
\frac{T_p}{2}}E\left\{{\left\vert{}r\left(t^{'}\right)\right\vert{}}^6\right\}dt^{'}=
\frac{Rk{\Gamma{}}^2}{T_p}\int_{-\ \frac{T_p}{2}}^{+\
\frac{T_p}{2}}E\left\{{\left\vert{}a(t^{'})+n\left(t^{'}\right)\right\vert{}}^6\right\}dt^{'}}\nonumber \\
&{=\frac{Rk{\Gamma{}}^2}{T_p}\int_{-\
\frac{T_p}{2}}^{+\frac{T_p}{2}}E\left\{{\left\vert{}a(t^{'})+p\left(t^{'}\right)+jq(t^{'})\right\vert{}}^6\right\}dt^{'}}
\end{align}
Defining $p_1(t^{'})\triangleq{}a(t^{'})+p\left(t^{'}\right)$ to
simplify mathematical expressions, \eqref{mom1_ebarate1_B} can be rewritten as below:

\begin{align}
\label{mom1_ebarate2_B}
&{{\mu{}}_1^{(1)}=\frac{Rk{\Gamma{}}^2}{T_p}\int_{-\ \frac{T_p}{2}}^{+\
\frac{T_p}{2}}E\left\{{\left\vert{}p_1\left(t^{'}\right)+jq\left(t^{'}\right)\right\vert{}}^6\right\}dt^{'}=\frac{Rk{\Gamma{}}^2}{T_p}\int_{-\
\frac{T_p}{2}}^{+\
\frac{T_p}{2}}E\left\{{\left(p_1^2\left(t^{'}\right)+q^2\left(t^{'}\right)\right)}^3\right\}dt^{'}} \nonumber \\
&{=\frac{Rk{\Gamma{}}^2}{T_p}\int_{-\ \frac{T_p}{2}}^{+\
\frac{T_p}{2}}{E\left\{p_1^6\left(t^{'}\right)+3p_1^4\left(t^{'}\right)q^2\left(t^{'}\right)+3p_1^2\left(t^{'}\right)q^4\left(t^{'}\right)+q^6\left(t^{'}\right)\right\}dt^{'}}}
\end{align}

Due to the statistical independence of $p_1(t^{'})$ with $q(t^{'})$ discussed
in section \ref{sec_system_math_model}, \eqref{mom1_ebarate2_B} can be simplified as:
\begin{align}
\label{mom1_ebarate3_B}
&{{\mu{}}_1^{(1)}=\frac{Rk{\Gamma{}}^2}{T_p}\int_{-\frac{T_p}{2}}^{+\frac{T_p}{2}}{\left({E\left\{p_1^6\left(t^{'}\right)\right\}+3E\left\{p_1^4\left(t^{'}\right)\right\}E\left\{q^2\left(t^{'}\right)\right\}}\right)dt^{'}}} \nonumber \\
&{+\frac{Rk{\Gamma{}}^2}{T_p}\int_{-\frac{T_p}{2}}^{+\frac{T_p}{2}}{\left({3E\left\{p_1^2\left(t^{'}\right)\right\}E\left\{q^4\left(t^{'}\right)\right\}+E\left\{q^6\left(t^{'}\right)\right\}}\right)dt^{'}}}
\end{align}
Using expressions for the arbitrary order moments of Gaussian random variable in \cite{winkelbauer2012moments} and noting that ${p_1{\left(t^{'}\right)}}\sim Normal{({a\left(t^{'}\right)},{\sigma_0}^2)}$ for all $t^{'}$ and ${q{\left(t^{'}\right)}}\sim Normal{\left(0,{\sigma_0}^2\right)}$ for all $t^{'}$ as discussed in section \ref{sec_system_math_model} of the paper, the following expressions are obtained:

\begin{align}
\label{gauss_moms_B}
&{E\left\{p_1^6\left(t^{'}\right)\right\}=a^6(t^{'})+15a^4(t^{'}){\sigma{}}_0^2+45a^2(t^{'}){\sigma{}}_0^4+15{\sigma{}}_0^6} \\ \nonumber
&{E\left\{p_1^4\left(t^{'}\right)\right\}=a^4(t^{'})+6a^2(t^{'}){\sigma{}}_0^2+3{\sigma{}}_0^4} \\ \nonumber
&{E\left\{p_1^2\left(t^{'}\right)\right\}=a^2(t^{'})+{\sigma{}}_0^2} \\ \nonumber
&{E\left\{q^6\left(t^{'}\right)\right\}=15{\sigma{}}_0^6} \\ \nonumber
&{E\left\{q^4\left(t^{'}\right)\right\}=3{\sigma{}}_0^4} \\ \nonumber
&{E\left\{q^2\left(t^{'}\right)\right\}={\sigma{}}_0^2}
\end{align}
Substituting expressions \eqref{gauss_moms_B} into \eqref{mom1_ebarate3_B} we have:

\begin{align}
\label{mom1_ebarate4_B}
&{{\mu{}}_1^{(1)}=\frac{Rk{\Gamma{}}^2}{T_p}\int_{-\ \frac{T_p}{2}}^{+\
\frac{T_p}{2}}\left(a^6(t^{'})+18{\sigma{}}_0^2a^4(t^{'})+72{\sigma{}}_0^4a^2(t^{'})+48{\sigma{}}_0^6\right)dt^{'}} \\ \nonumber
&{=\frac{Rk{\Gamma{}}^2}{T_p}\int_{-\
\frac{T_p}{2}}^{+\
\frac{T_p}{2}}\left(P_r^3 Sinc^6({t^{'}}/{{\tau{}}_c})+18{\sigma{}}_0^2P_r^2  Sinc^4({t^{'}}/{{\tau{}}_c})+72{\sigma{}}_0^4 P_r Sinc^2({t^{'}}/{{\tau{}}_c})+48{\sigma{}}_0^6\right)dt^{'}}
\end{align}
Changing the integral variable $u={t^{'}}/{{\tau{}}_c}$ it can be shown that:
\begin{align}
\label{mom1_ebarate5_B}
{\mu{}}_1^{(1)}=\frac{Rk{\Gamma{}}^2}{PRD}\int_{-\ \frac{PRD}{2}}^{+\
\frac{RRD}{2}}\left(P_r^3 Sinc^6(u)+18{\sigma{}}_0^2P_r^2 Sinc^4(u)+72{\sigma{}}_0^4P_r Sinc^2(u)+48{\sigma{}}_0^6\right)du
\end{align}
For obtaining \eqref{mom1_ebarate5_B} and for $PRD\gg{}1$, three integral forms must be calculated as below:
\begin{equation}
\label{sinc2_B}
\int_{-\ {PRD}/{2}}^{+\
{RRD}/{2}}Sinc^2(u)du\cong{}1
\end{equation} 
\begin{equation}
\label{sinc4_B}
\int_{-\ {PRD}/{2}}^{+\
{RRD}/{2}}Sinc^4(u)du\cong{}0.667
\end{equation} 
\begin{equation}
\label{sinc6_B}
\int_{-\ {PRD}/{2}}^{+\
{RRD}/{2}}Sinc^6(u)du\cong{}0.55
\end{equation} 
Finally, the mean of the decision variable can
be obtained as follows:
\begin{equation}
\label{mom1_b_1_final_B}
{\mu{}}_1^{(1)}=\frac{Rk{\Gamma{}}^2}{PRD}(48{\sigma{}}_0^6 PRD +72{\sigma{}}_0^4 P_r+12{\sigma{}}_0^2P_r^2+0.55 P_r^3)
\end{equation}
For $b=0$, the mean of the decision variable can be obtained using \eqref{mom1_b_1_final_B}
and holding $P_r=0$ as below:
\begin{equation}
\label{mom1_b_0_final_B}
{\mu{}}_1^{(0)}\triangleq{}E\left\{{Y}/{b=0}\right\}=48Rk{\Gamma{}}^2{\sigma{}}_0^6
\end{equation}
\section{Driving Second Order Moment\label{app_mom2}}
In this appendix we intend to drive a mathematical expression for the second order moment of the decision variable $Y$. Following the approach employed in appendix B, we begin by assuming that the transmitted bit is bit one ($b=1$). And also in simplifying mathematical expressions, the variance of the decision variable $Y$ is first calculated, then, the second order moment is obtained based on the mathematical expression of the variance, i.e., ${{\mu}_2}^{(b)}={Var}^{(b)}+{\left({\mu}_1^{(b)}\right)}^2$. The variance term can be expressed as follows:
\begin{align}
\label{Var_ebarate1_C}
&{{{Var}^{(1)}\triangleq{}E\left\{{Y^2}/{b=1}\right\}-E^2\left\{{Y}/{b=1}\right\}\triangleq{}\mu{}}_2^{(1)}-{\left({\mu{}}_1^{(1)}\right)}^2}\\ \nonumber
&{=\frac{R^2k^2{\Gamma{}}^4}{T_p^2}\int_{-\frac{T_p}{2}}^{\frac{T_p}{2}}\int_{-\frac{T_p}{2}}^{\frac{T_p}{2}}E\left\{{\left\vert{}r\left(t^{'}\right)\right\vert{}}^6{\left\vert{}r\left(t^{''}\right)\right\vert{}}^6\right\}\
dt^{'}dt^{''}}\\ \nonumber
&{-\
\frac{R^2k^2{\Gamma{}}^4}{T_p^2}\int_{-\frac{T_p}{2}}^{\frac{T_p}{2}}\int_{-\frac{T_p}{2}}^{\frac{T_p}{2}}E\left\{{\left\vert{}r\left(t^{'}\right)\right\vert{}}^6\right\}E\left\{{\left\vert{}r\left(t^{''}\right)\right\vert{}}^6\right\}\
dt^{'}dt^{''}}\\ \nonumber 
&{=\frac{R^2k^2{\Gamma{}}^4}{T_p^2}\int_{-\ \frac{T_p}{2}}^{+\
\frac{T_p}{2}}\int_{-\ \frac{T_p}{2}}^{+\
\frac{T_p}{2}}J(t^{'},t^{''})dt^{'}dt^{''}}
\end{align}
Where $J\left(t^{'},t^{''}\right)$ in \eqref{Var_ebarate1_C}, is expressed as:
\begin{equation}
\label{J_C}
J\left(t^{'},t^{''}\right)={J_1}\left(t^{'},t^{''}\right)-J_2\left(t^{'},t^{''}\right)
\end{equation}
Where the expressions of ${J_1}\left(t^{'},t^{''}\right)$ and ${J_2}\left(t^{'},t^{''}\right)$ are:
\begin{align}
\label{J1_C}
{{J_1}\left(t^{'},t^{''}\right)\ ={}E\binom{\left\{p_1^6\left(t^{'}\right)+3p_1^4\left(t^{'}\right)q^2\left(t^{'}\right)+3p_1^2\left(t^{'}\right)q^4\left(t^{'}\right)+q^6(t^{'})\right\}}{\times{}\left\{p_1^6\left(t^{''}\right)+3p_1^4\left(t^{''}\right)q^2\left(t^{''}\right)+3p_1^2\left(t^{''}\right)q^4\left(t^{''}\right)+q^6(t^{''})\right\}}}
\end{align}
\begin{align}
\label{J2_C}
{{J_2}\left(t^{'},t^{''}\right)\ ={}{\binom{E\left\{p_1^6\left(t^{'}\right)+3p_1^4\left(t^{'}\right)q^2\left(t^{'}\right)+3p_1^2\left(t^{'}\right)q^4\left(t^{'}\right)+q^6(t^{'})\right\}}{\times{}E\left\{p_1^6\left(t^{''}\right)+3p_1^4\left(t^{''}\right)q^2\left(t^{''}\right)+3p_1^2\left(t^{''}\right)q^4\left(t^{''}\right)+q^6(t^{''})\right\}}}}
\end{align}
Employing variable change of $u^{'}={t^{'}}/{{\tau{}}_c}$ and
$u^{''}={t^{''}}/{{\tau{}}_c}$, the form of the above expressions for the variance would be:
\begin{align}
\label{Var_ebarate2_C}
Var^{(1)}=\ \frac{R^2k^2{\Gamma{}}^4}{PRD^2}\int_{-\ \frac{PRD}{2}}^{+\
\frac{PRD}{2}}\int_{-\ \frac{PRD}{2}}^{+\
\frac{PRD}{2}}\left(J_1\left(u^{'},u^{''}\right)-J_2\left(u^{'},u^{''}\right)\right)\
du^{'}du^{''}\end{align}
\begin{align}
\label{J1_riz_shode_C}
J_1\left(u^{'},u^{''}\right)=E\binom{\left\{p_1^6\left(u^{'}{\tau{}}_c\right)+3p_1^4\left(u^{'}{\tau{}}_c\right)q^2\left(u^{'}{\tau{}}_c\right)+3p_1^2\left(u^{'}{\tau{}}_c\right)q^4\left(u^{'}{\tau{}}_c\right)+q^6(u^{'}{\tau{}}_c)\right\}}{\times{}\left\{p_1^6\left(u^{''}{\tau{}}_c\right)+3p_1^4\left(u^{''}{\tau{}}_c\right)q^2\left(u^{''}{\tau{}}_c\right)+3p_1^2\left(u^{''}{\tau{}}_c\right)q^4\left(u^{''}{\tau{}}_c\right)+q^6(u^{''}{\tau{}}_c)\right\}}
\end{align}
\begin{align}
\label{J2_riz_shode_C}
J_2\left(u^{'},u^{''}\right)=\binom{E\left\{p_1^6\left(u^{'}{\tau{}}_c\right)+3p_1^4\left(u^{'}{\tau{}}_c\right)q^2\left(u^{'}{\tau{}}_c\right)+3p_1^2\left(u^{'}{\tau{}}_c\right)q^4\left(u^{'}{\tau{}}_c\right)+q^6(u^{'}{\tau{}}_c)\right\}}{\times{}E\left\{p_1^6\left(u^{''}{\tau{}}_c\right)+3p_1^4\left(u^{''}{\tau{}}_c\right)q^2\left(u^{''}{\tau{}}_c\right)+3p_1^2\left(u^{''}{\tau{}}_c\right)q^4\left(u^{''}{\tau{}}_c\right)+q^6(u^{''}{\tau{}}_c)\right\}}
\end{align}
We note that in the integrand function in \eqref{Var_ebarate2_C}, four random variables exist namely, $p_1(u^{'}{\tau{}}_c)\triangleq{}V_1$,
$p_1(u^{''}{\tau{}}_c)\triangleq{}V_2$, $q_1(u^{'}{\tau{}}_c)\triangleq{}V_3$ and $q_1(u^{''}{\tau{}}_c)\triangleq{}V_4$ which in accordance to our discussions in section \ref{sec_system_math_model} of the paper, are jointly Gaussian
random variables with mean vector and covariance matrix as:
\begin{equation}
\label{V_vector_C}
\vec{V}={\left[\begin{array}{
cccc}
V_1 & V_2 & V_3 & V_4
\end{array}\right]}^T
\end{equation}
\begin{equation}
\label{Mean_vector_C}
Mean\ Vector\triangleq{}m=\sqrt{P_r}{\left[\begin{array}{
cccc}
a_1 & a_2 & 0 & 0
\end{array}\right]}^T
\end{equation}
\begin{equation}
\label{Cov_Matrix_C}
Cov.\ Matrix\triangleq{}C={\sigma{}}_0^2\times{}\left[\begin{array}{
cccc}
1 & R_1 & 0 & 0 \\
R_1 & 1 & 0 & 0 \\
0 & 0 & 1 & R_1 \\
0 & 0 & R_1 & 1
\end{array}\right]
\end{equation}
where in the above equations $a_1\triangleq{}{a(u^{'})}/{\sqrt{P_r}}\triangleq{}Sinc\left(u^{'}\right)$,
$a_2\triangleq{}{a(u^{''})}/{\sqrt{P_r}}\triangleq{}Sinc\left(u^{''}\right)$ and $R_1\triangleq{}sinc(u^{'}-u^{''})$. Since the elements of the vector presented in \eqref{V_vector_C} are jointly normal, the joint characteristic function of the decision random
variable can be expressed as follows:
\begin{equation}
\label{S_vector_C}
\vec{S}={\left[\begin{array}{
cccc}
s_1 & s_2 & s_3 & s_4
\end{array}\right]}^T
\end{equation}
\begin{equation}
\label{char_func_C}
\psi{}\left(s_1,s_2,s_3,s_4\right)\triangleq{}E\left\{e^{s_1V_1+s_2V_2+s_3V_3+s_4V_4}\right\}=E\left\{e^{{{\vec{S}}}^T\times{}m+\frac{1}{2}{\
{\vec{S}}^{T}\times{}C\times{}{\vec{S}}}}\right\}
\end{equation}
The joint moments of random variables presented in \eqref{V_vector_C} can be expressed using the joint characteristic function as follows:
\begin{equation}
\label{joint_moments_v1_ta_v4_C}
E\left\{V_1^{n_1}V_2^{n_2}V_3^{n_3}V_4^{n_4}\right\}={\left.\frac{{\partial{}}^{\left(n_1+n_2+n_3+n_4\right)}\psi{}}{\partial{}s_1^{n_1}\partial{}s_2^{n_2}\partial{}s_3^{n_3}\partial{}s_4^{n_4}}\right\vert{}}_{\begin{array}{
cccc}
s_i=0 ; i=1,2,3,4 \
\end{array}}
\end{equation}
Therefore, \eqref{J1_riz_shode_C} and \eqref{J2_riz_shode_C} can be readily calculated with the help
of joint characteristic function as follows:
\begin{equation}
\label{J_1_barhasbe_u_uprim_C}
J_1(u^{'},u^{''})={\sum_{n=1}^4\sum_{m=1}^4\left.\frac{z_{1n}z_{1m\
}{\partial{}}^{12}\psi{}}{\partial{}s_1^{z_{2n}}\partial{}s_2^{z_{2m}}\partial{}s_3^{z_{3n}}\partial{}s_4^{z_{3m}}}\right\vert{}}_{\begin{array}{
cccc}
s_i=0 ; i=1,2,3,4 \
\end{array}}
\end{equation}
\begin{equation}
\label{J_2_barhasbe_u_uprim_C}
J_2(u^{'},u^{''})=\left({\left.\sum_{n=1}^4\frac{z_{1n}{\partial{}}^6\psi{}}{\partial{}s_1^{z_{2n}}\partial{}s_3^{z_{3n}}}\right\vert{}}_{\begin{array}{
cccc}
s_i=0 \
\end{array}}\right)\times{}\left({\left.\sum_{m=1}^4\frac{z_{1m}{\partial{}}^6\psi{}}{\partial{}s_2^{z_{2m}}\partial{}s_4^{z_{3m}}}\right\vert{}}_{\begin{array}{
cccc}
s_i=0 \
\end{array}}\right) 
\end{equation}
Which $z_{ij}$ in \eqref{J_1_barhasbe_u_uprim_C} and \eqref{J_2_barhasbe_u_uprim_C} is:
\begin{equation}
\label{Z_matrix_C}
\
Z=\left[z_{ij}\right]=\left[\begin{array}{
cccc}
1 & 3 & 3 & 1 \\
6 & 4 & 2 & 0 \\
0 & 2 & 4 & 6
\end{array}\right] \
\end{equation}
Using the symbolic tools of Matlab software, $J_1$ and $J_2$ in \eqref{J_1_barhasbe_u_uprim_C} and \eqref{J_2_barhasbe_u_uprim_C} are obtained and then replaced in \eqref{Var_ebarate2_C}, the variance can be written as:
\begin{equation}
\label{var1_sigma_2_integ_C}
Var^{(1)}=\frac{R^2k^2{\Gamma{}}^4}{PRD^2}\sum_{i=1}^{28}r_i{\sigma{}}_0^{g_i}P_r^{n_i} I_{0,i}
\end{equation} 
Where $I_{0,i}$ in \eqref{var1_sigma_2_integ_C} can be expressed as follows:
\begin{equation}
\label{I_0_i_C}
I_{0,i}=\left(\int_{-\
{PRD}/{2}}^{+\ {PRD}/{2}}\int_{-\ {PRD}/{2}}^{+\
{PRD}/{2}}R_1^{h_i}a_1^{l_i}a_2^{x_i}\ du^{'}du^{''}\right)\
\end{equation}
Where $r_i$, $g_i$, $\ h_i$, $l_i$, $n_i$ and $x_i$  are constant coefficients obtained from symbolic calculations and are shown in \tablename{ \ref{table_I1_C}}. Using values in \tablename{ \ref{table_I1_C}}, summing all terms in \eqref{var1_sigma_2_integ_C}, the variance can be expressed as:
\begin{table}[H]
\caption{Terms in the equation \eqref{var1_sigma_2_integ_C}} 
\centering 
\begin{tabular}{c c c c c c c c} 
\hline\hline 
$i$ & $r_i$ & $g_i$ & $n_i$ & $h_i$ & $l_i$ & $x_i$ & $I_{0,i}$ \\ 
\hline 
1 &	2304 &	12 &	0 &	6 &	0 &	0 &	$0.55\times PRD$ \\
2 &	6912 &	10 &	1 &	5 &	1 &	1 &	$0.52$ \\
3 &	6912 &	8 &	2 &	4 &	2 &	2 &	$0.394$ \\
4 &	10368 &	10 &	1 &	4 &	2 &	0 &	$0.64$ \\
5 &	10368 &	10 &	1 &	4 &	0 &	2 &	$0.64$ \\
6 &	20736 &	12 &	0 &	4 &	0 &	0 &	$0.667\times PRD$ \\
7 &	2688 &	6 &	3 &	3 &	3 &	3 &	$0.343$ \\
8 &	10368 &	8 &	2 &	3 &	3 &	1 &	$0.45$ \\
9 &	10368 &	8 &	2 &	3 &	1 &	3 &	$0.45$ \\
10 &	41472 &	10 &	1 &	3 &	1 &	1 &	$0.657$ \\
11 &	468 &	4 &	4 &	2 &	4 &	4 &	$0.317$ \\
12 &	3456 &	6 &	3 &	2 &	4 &	2 &	$0.395$ \\
13 &	2592 &	8 &	2 &	2 &	4 &	0 &	$0.66$ \\
14 &	3456 &	6 &	3 &	2 &	2 &	4 &	$0.3945$ \\
15 &	25920 &	8 &	2 &	2 &	2 &	2 &	$0.5$ \\
16 &	20736 &	10 &	1 &	2 &	2 &	0 &	$1$ \\
17 &	2592 &	8 &	2 &	2 &	0 &	4 &	$0.665$ \\
18 &	20736 &	10 &	1 &	2 &	0 &	2 &	$1$ \\
19 &	20736 &	12 &	0 &	2 &	0 &	0 &	$1\times PRD$ \\
20 &	36 &	2 &	5 &	1 &	5 &	5 &	$0.3043$ \\
21 &	432 &	4 &	4 &	1 &	5 &	3 &	$0.37$ \\
22 &	864 &	6 &	3 &	1 &	5 &	1 &	$0.55$ \\
23 &	432 &	4 &	4 &	1 &	3 &	5 &	$0.37$ \\
24 &	5184 &	6 &	3 &	1 &	3 &	3 &	$0.45$ \\
25 &	10368 &	8 &	2 &	1 &	3 &	1 &	$0.66$ \\
26 &	864 &	6 &	3 &	1 &	1 &	5 &	$0.55$ \\
27 &	10368 &	8 &	2 &	1 &	1 &	3 &	$0.66$ \\
28 &	20736 &	10 &	1 &	1 &	1 &	1 &	$1$ \\ [1ex]

\hline 
\end{tabular}
\label{table_I1_C} 
\end{table}

\begin{equation}
\label{var_b_1_final_C}
Var^{(1)}=\frac{R^2k^2{\Gamma{}}^4}{PRD^2}\times{}\left(\begin{array}{l}\
35843{\sigma{}}_0^{12}\times{}PRD+106320{\sigma{}}_0^{10}P_r+42135{\sigma{}}_0^8P_r^2
\\
+6933.7{\sigma{}}_0^6P_r^3+468{\sigma{}}_0^4P_r^4+10.95{\sigma{}}_0^2P_r^5\
\end{array}\right)
\end{equation}
Also the variance for transmitting bit zero ($b=0$) can be easily derived from \eqref{var_b_1_final_C}. Equating  $P_r=0$, the variance for bit zero is:
\begin{equation}
\label{var_b_0_final_C}
Var^{(0)}=\frac{35843R^2k^2{\Gamma{}}^4{\sigma{}}_0^{12}}{PRD}
\end{equation}
Using the mean values (${\mu{}}_1^{(1)}$,${\mu{}}_1^{(0)}$) from appendix \ref{app_mom1}
 and variances from \eqref{var_b_1_final_C} and \eqref{var_b_0_final_C}, second order moments for the bit one and the bit zero can be expressed as follows:
\begin{equation}
\label{mom2_final_b_1_C}
{\mu{}}_2^{(1)}=\
\frac{R^2k^2{\Gamma{}}^4}{PRD^2}\times{}\left(\begin{array}{l}2304{\sigma{}}_0^{12}\times{}PRD^2+35834{\sigma{}}_0^{12}\times{}PRD+6912{\sigma{}}_0^{10}P_r\times{}PRD
\\
+1152{\sigma{}}_0^8P_r^2\times{}PRD+52.8{\sigma{}}_0^6P_r^3\times{}PRD+106320{\sigma{}}_0^{10}P_r
\\
+47319{\sigma{}}_0^8P_r^2+8661.7{\sigma{}}_0^6P_r^{3\
}+691.2{\sigma{}}_0^4P_r^4+24.15{\sigma{}}_0^2P_r^5 \\
+0.3025P_r^6\end{array}\right)
\end{equation}
\begin{equation}
\label{mom2_final_b_0_C}
{\mu{}}_2^{(0)}=\frac{R^2k^2{\Gamma{}}^4}{PRD^2}\left(2304{\sigma{}}_0^{12}\times{}PRD^2+35834{\sigma{}}_0^{12}\times{}PRD\right)
\end{equation}
\section{Driving Third Order Moment \label{app_mom3}}
In this appendix the third order moment of the decision variable is derived. Similar to the approach employed in the previous appendix, the moment is calculated for transmitted bit one ($b=1$), after that the moment is calculated for $b=0$. The third order moment can be expressed as:
\begin{equation}
\label{ebarate_integralie_mom3_D}
{{\mu{}}_3^{\left(1\right)}=\frac{R^3k^3{\Gamma{}}^6}{PRD^3}\times{}\int_{-
{PRD}/{2}}^{+{PRD}/{2}}\int_{-{PRD}/{2}}^{+
{PRD}/{2}}\int_{-{PRD}/{2}}^{+
{PRD}/{2}}E\left\{G\left(u,u^{'},u^{''}\right)\ \right\}du\ du^{'}du^{''}}
\end{equation}
Where in \eqref{ebarate_integralie_mom3_D} $G\left(u,u^{'},u^{''}\right)=G_0\left(u\right)G_0(u^{'})G_0(u^{''})$. Also $G_0(x)$ can be expressed as:
\begin{equation}
\label{G_0_D}
G_0\left(x\right)=\left\{\begin{array}{l}p_1^6\left(x{\tau{}}_c\right)+3p_1^4\left(x{\tau{}}_c\right)q^2\left(x{\tau{}}_c\right)
\\
+3p_1^2\left(x{\tau{}}_c\right)q^4\left(x{\tau{}}_c\right)+q^6\left(x{\tau{}}_c\right)\end{array}\right\}
\end{equation}
Similar to the approach employed in appendix \ref{app_mom2}, with using the joint characteristic function of six jointly normal random variables namely,
$p_1(u{\tau{}}_c)\triangleq{}V_1$, $p_1(u^{'}{\tau{}}_c)\triangleq{}V_2$,
$p_1(u^{''}{\tau{}}_c)\triangleq{}V_3$, $q(u{\tau{}}_c)\triangleq{}V_4$,
$q(u^{'}{\tau{}}_c)\triangleq{}V_5$ and $q(u^{''}{\tau{}}_c)\triangleq{}V_6$, now by
defining:
\begin{equation}
\label{rv_vector_V_D}
{V}={\left[\begin{array}{
cccccc}
V_1 & V_2 & V_3 & V_4 & V_5 & V_6
\end{array}\right]}^T
\end{equation}
\begin{equation}
\label{mean_vector_D}
Mean\ Vector\ \triangleq{}m=\sqrt{P_r}\times{}\ {\left[\begin{array}{
cccccc}
a_1 & a_2 & a_3 & 0 & 0 & 0
\end{array}\right]}^T
\end{equation}
\begin{equation}
\label{cov_matrix_D}
Cov.\ Matrix\triangleq{}C={\sigma{}}_0^2\times{}\left[\begin{array}{
cccccc}
1 & R_1 & R_2 & 0 & 0 & 0 \\
R_1 & 1 & R_3 & 0 & 0 & 0 \\
R_2 & R_3 & 1 & 0 & 0 & 0 \\
0 & 0 & 0 & 1 & R_1 & R_2 \\
0 & 0 & 0 & R_1 & 1 & R_3 \\
0 & 0 & 0 & R_2 & R_3 & 1
\end{array}\right]
\end{equation}
\begin{equation}
\label{S_vector_D}
{\vec{S}}={\left[\begin{array}{
cccccc}
s_1 & s_2 & s_3 & s_4 & s_5 & s_6
\end{array}\right]}^T
\end{equation}
\begin{equation}
\label{Char_func_tarif_D}
\psi{}\left(s_1,s_2,s_3,s_4,s_5,s_6\right)\triangleq{}E\left\{e^{\sum_{i=1}^6s_iV_i}\right\}=E\left\{e^{{{\vec{S}}}^T\times{}m+\frac{1}{2}{\
{\vec{S}}^{T}\times{}C\times{}{\vec{S}}}}\right\}
\end{equation}
Which in \eqref{mean_vector_D}, $a_1\triangleq{}\frac{a(u)}{\sqrt{P_r}}=Sinc\left(u\right)$, 
$a_2\triangleq{}\frac{a(u^{'})}{\sqrt{P_r}}=Sinc\left(u^{'}\right)$and
$a_3\triangleq{}\frac{a(u^{''})}{\sqrt{P_r}}=Sinc\left(u^{''}\right)$ and in \eqref{cov_matrix_D}, $R_1\triangleq{}Sinc\left(u-u^{'}\right)$, 
$R_2\triangleq{}Sinc\left(u-u^{''}\right)$ and $R_3\triangleq{}Sinc\left(u^{'}-u^{''}\right)$. $E\left\{G\left(u,u^{'},u^{''}\right)\right\}$ appeared in \eqref{ebarate_integralie_mom3_D}, can be expressed as:
\begin{equation}
\label{G_barhasbe_char_func_D}
{E\left\{G\left(u,u^{'},u^{''}\right)\right\}=\sum_{n=1}^4\sum_{m=1}^4\left.\sum_{k=1}^4\frac{z_{1n}z_{1m}z_{1k}{\partial{}}^{18}\psi{}}{\partial{}s_1^{z_{2n}}\partial{}s_2^{z_{2m}}\partial{}s_3^{z_{2k}}\partial{}s_4^{z_{3n}}\partial{}s_5^{z_{3m}}\partial{}s_6^{z_{3k}}}\right\vert{}}_{s_i=0}
\end{equation}
In which $z_{ij}$s are given in equation \eqref{Z_matrix_C} in the previous appendix. \eqref{G_barhasbe_char_func_D} can be obtained using symbolic tools of Matlab software and the result contains sum of polynomials in the form $R_1^{n_1}R_2^{n_2}R_3^{n_3}a_1^{m_1}a_2^{m_2}a_3^{m_3}$. These terms can be classified in the cases below.
\subsection{Case $n_1=0,\ n_2=0, \ n_3=0,\  m_1=0, \ m_2=0, \ m_3=0$}
$110592{\sigma{}}_0^{18}$

\subsection{Case $n_1>0,\ n_2=0, \ n_3=0,\  m_1=0, \ m_2=0, \ m_3=0$}
$110592R_1^6{\sigma{}}_0^{18}+995328R_1^4{\sigma{}}_0^{18}+995328R_1^2{\sigma{}}_0^{18}$

\subsection{Case $n_1=0,\ n_2>0, \ n_3=0,\  m_1=0, \ m_2=0, \ m_3=0$}
$110592R_2^6{\sigma{}}_0^{18}+995328R_2^4{\sigma{}}_0^{18}+995328R_2^2{\sigma{}}_0^{18}$

\subsection{Case $n_1=0,\ n_2=0, \ n_3>0,\  m_1=0, \ m_2=0, \ m_3=0$}
$110592R_3^6{\sigma{}}_0^{18}+995328R_3^4{\sigma{}}_0^{18}+995328R_3^2{\sigma{}}_0^{18}$

\subsection{Case $n_1=0,\ n_2=0, \ n_3=0,\  m_1>0, \ m_2=0, \ m_3=0$}
$2304P_r^3{\sigma{}}_0^{12}a_1^6+41472P_r^2{\sigma{}}_0^{14}a_1^4+165888{P_r{\sigma{}}_0^{16}a}_1^2$

\subsection{Case $n_1=0,\ n_2=0, \ n_3=0,\  m_1=0, \ m_2>0, \ m_3=0$}
$2304P_r^3{\sigma{}}_0^{12}a_2^6+41472P_r^2{\sigma{}}_0^{14}a_2^4+165888{P_r{\sigma{}}_0^{16}a}_2^2$

\subsection{Case $n_1=0,\ n_2=0, \ n_3=0,\  m_1=0, \ m_2=0, \ m_3>0$}
$2304P_r^3{\sigma{}}_0^{12}a_3^6+41472P_r^2{\sigma{}}_0^{14}a_3^4+165888{P_r{\sigma{}}_0^{16}a}_3^2$

\subsection{Case $n_1>0,\ n_2>0, \ n_3=0,\  m_1=0, \ m_2=0, \ m_3=0$}
$2985984R_1^4R_2^2{\sigma{}}_0^{18}+2985984R_1^2R_2^4{\sigma{}}_0^{18}+5971968R_1^2R_2^2{\sigma{}}_0^{18}$

\subsection{Case $n_1>0,\ n_2=0, \ n_3>0,\  m_1=0, \ m_2=0, \ m_3=0$}
$2985984R_1^4R_3^2{\sigma{}}_0^{18}+2985984R_1^2R_3^4{\sigma{}}_0^{18}+5971968R_1^2R_3^2{\sigma{}}_0^{18}$

\subsection{Case $n_1=0,\ n_2>0, \ n_3>0,\  m_1=0, \ m_2=0, \ m_3=0$}
$2985984R_2^4R_3^2{\sigma{}}_0^{18}+2985984R_2^2R_3^4{\sigma{}}_0^{18}+5971968R_2^2R_3^2{\sigma{}}_0^{18}$

\subsection{Case $n_1>0,\ n_2>0, \ n_3>0,\  m_1=0, \ m_2=0, \ m_3=0$}
$1990656R_1^5R_2R_3{\sigma{}}_0^{18}+7962624R_1^4R_2^2R_3^2{\sigma{}}_0^{18}+6193152R_1^3R_2^3R_3^3{\sigma{}}_0^{18}
+11943936R_1^3R_2^3R_3{\sigma{}}_0^{18}+ \ 11943936R_1^3R_2R_3^3{\sigma{}}_0^{18}
+
11943936R_1^3R_2R_3{\sigma{}}_0^{18}+7962624R_1^2R_2^4R_3^2{\sigma{}}_0^{18}+7962624R_1^2R_2^2R_3^4{\sigma{}}_0^{18}+ \ 29859840R_1^2R_2^2R_3^2{\sigma{}}_0^{18}+1990656R_1R_2^5R_3{\sigma{}}_0^{18}+11943936R_1R_2^3R_3^3{\sigma{}}_0^{18}
+11943936R_1R_2^3R_3{\sigma{}}_0^{18}+ \ 1990656R_1R_2R_3^5{\sigma{}}_0^{18}+ 11943936R_1R_2R_3^3{\sigma{}}_0^{18}+5971968R_1R_2R_3{\sigma{}}_0^{18}$

\subsection{Case $n_1>0,\ n_2=0, \ n_3=0,\  m_1>0, \ m_2=0, \ m_3=0$}
$497664R_1^4a_1^2P_r{\sigma{}}_0^{16}+124416R_1^2a_1^4P_r^2{\sigma{}}_0^{14}+995328R_1^2a_1^2P_r{\sigma{}}_0^{16}$
\subsection{Case $n_1=0,\ n_2>0, \ n_3=0,\  m_1>0, \ m_2=0, \ m_3=0$}
$497664R_2^4a_1^2P_r{\sigma{}}_0^{16}+124416R_2^2a_1^4P_r^2{\sigma{}}_0^{14}+995328R_2^2a_1^2P_r{\sigma{}}_0^{16}$

\subsection{Case $n_1=0,\ n_2=0, \ n_3>0,\  m_1>0, \ m_2=0, \ m_3=0$}
$2304R_3^6a_1^6P_r^3{\sigma{}}_0^{12}+41472R_3^6a_1^4P_r^2{\sigma{}}_0^{14}+165888R_3^6a_1^2P_r{\sigma{}}_0^{16}+20736R_3^4a_1^6P_r^3{\sigma{}}_0^{12}+\ 373248R_3^4a_1^4P_r^2{\sigma{}}_0^{14}+1492992R_3^4a_1^2P_r{\sigma{}}_0^{16}+20736R_3^2a_1^6P_r^3{\sigma{}}_0^{12}+373248R_3^2a_1^4P_r^2{\sigma{}}_0^{14}+ \ 1492992R_3^2a_1^2P_r{\sigma{}}_0^{16}$

\subsection{Case $n_1>0,\ n_2=0, \ n_3=0,\  m_1=0, \ m_2>0, \ m_3=0$}
$497664R_1^4a_2^2P_r{\sigma{}}_0^{16}+124416R_1^2a_2^4P_r^2{\sigma{}}_0^{14}+995328R_1^2a_2^2P_r{\sigma{}}_0^{16}$

\subsection{Case $n_1=0,\ n_2>0, \ n_3=0,\  m_1=0, \ m_2>0, \ m_3=0$}
$2304R_2^6a_2^6P_r^3{\sigma{}}_0^{12}+41472R_2^6a_2^4P_r^2{\sigma{}}_0^{14}+165888R_2^6a_2^2P_r{\sigma{}}_0^{16}+20736R_2^4a_2^6P_r^3{\sigma{}}_0^{12}+\ 373248R_2^4a_2^4P_r^2{\sigma{}}_0^{14}+1492992R_2^4a_2^2P_r{\sigma{}}_0^{16}+20736R_2^2a_2^6P_r^3{\sigma{}}_0^{12}+373248R_2^2a_2^4P_r^2{\sigma{}}_0^{14}+ \ 
1492992R_2^2a_2^2P_r{\sigma{}}_0^{16}$

\subsection{Case $n_1=0,\ n_2=0, \ n_3>0,\  m_1=0, \ m_2>0, \ m_3=0$}
$497664R_3^4a_2^2P_r{\sigma{}}_0^{16}+124416R_3^2a_2^4P_r^2{\sigma{}}_0^{14}+995328R_3^2a_2^2P_r{\sigma{}}_0^{16}$

\subsection{Case $n_1>0,\ n_2=0, \ n_3=0,\  m_1=0, \ m_2=0, \ m_3>0$}
$2304R_1^6a_3^6P_r^3{\sigma{}}_0^{12}+41472R_1^6a_3^4P_r^2{\sigma{}}_0^{14}+165888R_1^6a_3^2P_r{\sigma{}}_0^{16}+20736R_1^4a_3^6P_r^3{\sigma{}}_0^{12}+ \ 373248R_1^4a_3^4P_r^2{\sigma{}}_0^{14}+1492992R_1^4a_3^2P_r{\sigma{}}_0^{16}+20736R_1^2a_3^6P_r^3{\sigma{}}_0^{12}+373248R_1^2a_3^4P_r^2{\sigma{}}_0^{14}+ \ 
1492992R_1^2a_3^2P_r{\sigma{}}_0^{16}$

\subsection{Case $n_1=0,\ n_2>0, \ n_3=0,\  m_1=0, \ m_2=0, \ m_3>0$}
$497664R_2^4a_3^2P_r{\sigma{}}_0^{16}+124416R_2^2a_3^4P_r^2{\sigma{}}_0^{14}+995328R_2^2a_3^2P_r{\sigma{}}_0^{16}$

\subsection{Case $n_1=0,\ n_2=0, \ n_3>0,\  m_1=0, \ m_2=0, \ m_3>0$}
$497664R_2^4a_3^2P_r{\sigma{}}_0^{16}+124416R_2^2a_3^4P_r^2{\sigma{}}_0^{14}+995328R_2^2a_3^2P_r{\sigma{}}_0^{16}$

\subsection{Case $n_1>0,\ n_2=0, \ n_3=0,\  m_1>0, \ m_2>0, \ m_3=0$}
$331776R_1^5a_1a_2P_r{\sigma{}}_0^{16}+331776R_1^4a_1^2a_2^2P_r^2{\sigma{}}_0^{14}+129024R_1^3a_1^3a_2^3P_r^3{\sigma{}}_0^{12}+497664R_1^3a_1^3a_2P_r^2{\sigma{}}_0^{14}+ \\ 497664R_1^3a_1a_2^3P_r^2{\sigma{}}_0^{14}+1990656R_1^3a_1a_2P_r{\sigma{}}_0^{16}+22464R_1^2a_1^4a_2^4P_r^4{\sigma{}}_0^{10}
+165888R_1^2a_1^4a_2^2P_r^3{\sigma{}}_0^{12}+ \\ 165888R_1^2a_1^2a_2^4P_r^3{\sigma{}}_0^{12}+1244160R_1^2a_1^2a_2^2P_r^2{\sigma{}}_0^{14}+1728R_1a_1^5a_2^5P_r^5{\sigma{}}_0^8+20736R_1a_1^5a_2^3P_r^4{\sigma{}}_0^{10}+ \\ 41472R_1a_1^5a_2P_r^3{\sigma{}}_0^{12}+20736R_1a_1^3a_2^5P_r^4{\sigma{}}_0^{10}+248832R_1a_1^3a_2^3P_r^3{\sigma{}}_0^{12}+497664R_1a_1^3a_2P_r^2{\sigma{}}_0^{14}+ \\ 
41472R_1a_1a_2^5P_r^3{\sigma{}}_0^{12}+497664R_1a_1a_2^3P_r^2{\sigma{}}_0^{14}+995328R_1a_1a_2P_r{\sigma{}}_0^{16}$

\subsection{Case $n_1=0,\ n_2>0, \ n_3=0,\  m_1>0, \ m_2=0, \ m_3>0$}
$331776R_2^5a_1a_3P_r{\sigma{}}_0^{16}+331776R_2^4a_1^2a_3^2P_r^2{\sigma{}}_0^{14}+129024R_2^3a_1^3a_3^3P_r^3{\sigma{}}_0^{12}+497664R_2^3a_1^3a_3P_r^2{\sigma{}}_0^{14}+ \\ 497664R_2^3a_1a_3^3P_r^2{\sigma{}}_0^{14}+1990656R_2^3a_1a_3P_r{\sigma{}}_0^{16}+22464R_2^2a_1^4a_3^4P_r^4{\sigma{}}_0^{10}
+165888R_2^2a_1^4a_3^2P_r^3{\sigma{}}_0^{12}+ \\ 165888R_2^2a_1^2a_3^4P_r^3{\sigma{}}_0^{12}+1244160R_2^2a_1^2a_3^2P_r^2{\sigma{}}_0^{14}+1728R_2a_1^5a_3^5P_r^5{\sigma{}}_0^8+20736R_2a_1^5a_3^3P_r^4{\sigma{}}_0^{10}+ \\ 41472R_2a_1^5a_3P_r^3{\sigma{}}_0^{12}+20736R_2a_1^3a_3^5P_r^4{\sigma{}}_0^{10}+248832R_2a_1^3a_3^3P_r^3{\sigma{}}_0^{12}+497664R_2a_1^3a_3P_r^2{\sigma{}}_0^{14}+ \\ 
41472R_2a_1a_3^5P_r^3{\sigma{}}_0^{12}+497664R_2a_1a_3^3P_r^2{\sigma{}}_0^{14}+995328R_2a_1a_3P_r{\sigma{}}_0^{16}$

\subsection{Case $n_1=0,\ n_2=0, \ n_3>0,\  m_1=0, \ m_2>0, \ m_3>0$}
$331776R_3^5a_2a_3P_r{\sigma{}}_0^{16}+331776R_3^4a_2^2a_3^2P_r^2{\sigma{}}_0^{14}+129024R_3^3a_2^3a_3^3P_r^3{\sigma{}}_0^{12}+497664R_3^3a_2^3a_3P_r^2{\sigma{}}_0^{14}+ \\ 497664R_3^3a_2a_3^3P_r^2{\sigma{}}_0^{14}+1990656R_3^3a_2a_3P_r{\sigma{}}_0^{16}+22464R_3^2a_2^4a_3^4P_r^4{\sigma{}}_0^{10}
+165888R_3^2a_2^4a_3^2P_r^3{\sigma{}}_0^{12}+ \\ 165888R_3^2a_2^2a_3^4P_r^3{\sigma{}}_0^{12}+1244160R_3^2a_2^2a_3^2P_r^2{\sigma{}}_0^{14}+1728R_3a_2^5a_3^5P_r^5{\sigma{}}_0^8+20736R_3a_2^5a_3^3P_r^4{\sigma{}}_0^{10}+ \\ 41472R_3a_2^5a_3P_r^3{\sigma{}}_0^{12}+20736R_3a_2^3a_3^5P_r^4{\sigma{}}_0^{10}+248832R_3a_2^3a_3^3P_r^3{\sigma{}}_0^{12}+497664R_3a_2^3a_3P_r^2{\sigma{}}_0^{14}+ \\ 
41472R_3a_2a_3^5P_r^3{\sigma{}}_0^{12}+497664R_3a_2a_3^3P_r^2{\sigma{}}_0^{14}+995328R_3a_2a_3P_r{\sigma{}}_0^{16}$

\subsection{Case $n_1=0,\ n_2=0, \ n_3=0,\  m_1>0, \ m_2>0, \ m_3=0$}
$48a_1^6a_2^6P_r^6{\sigma{}}_0^6+864a_1^6a_2^4P_r^5{\sigma{}}_0^8+3456a_1^6a_2^2P_r^4{\sigma{}}_0^{10}+864a_1^4a_2^6P_r^5{\sigma{}}_0^8+ \ 15552a_1^4a_2^4P_r^4{\sigma{}}_0^{10}+62208a_1^4a_2^2P_r^3{\sigma{}}_0^{12}
+3456a_1^2a_2^6P_r^4{\sigma{}}_0^{10}+62208a_1^2a_2^4P_r^3{\sigma{}}_0^{12}+ \ 
248832a_1^2a_2^2P_r^2{\sigma{}}_0^{14}$

\subsection{Case $n_1=0,\ n_2=0, \ n_3=0,\  m_1>0, \ m_2=0, \ m_3>0$}
$48a_1^6a_3^6P_r^6{\sigma{}}_0^6+864a_1^6a_3^4P_r^5{\sigma{}}_0^8+3456a_1^6a_3^2P_r^4{\sigma{}}_0^{10}+864a_1^4a_3^6P_r^5{\sigma{}}_0^8+ \ 15552a_1^4a_3^4P_r^4{\sigma{}}_0^{10}+62208a_1^4a_3^2P_r^3{\sigma{}}_0^{12}
+3456a_1^2a_3^6P_r^4{\sigma{}}_0^{10}+62208a_1^2a_3^4P_r^3{\sigma{}}_0^{12}+ \ 
248832a_1^2a_3^2P_r^2{\sigma{}}_0^{14}$

\subsection{Case $n_1=0,\ n_2=0, \ n_3=0,\  m_1=0, \ m_2>0, \ m_3>0$}
$48a_2^6a_3^6P_r^6{\sigma{}}_0^6+864a_2^6a_3^4P_r^5{\sigma{}}_0^8+3456a_2^6a_3^2P_r^4{\sigma{}}_0^{10}+864a_2^4a_3^6P_r^5{\sigma{}}_0^8+$  $15552a_2^4a_3^4P_r^4{\sigma{}}_0^{10}+62208a_2^4a_3^2P_r^3{\sigma{}}_0^{12}
+3456a_2^2a_3^6P_r^4{\sigma{}}_0^{10}+62208a_2^2a_3^4P_r^3{\sigma{}}_0^{12}+ 
248832a_2^2a_3^2P_r^2{\sigma{}}_0^{14}$

\begin{table}[ht]
\caption{Integral results in the form of $I_1$ (equation \eqref{I1_D}) } 
\centering 
\begin{tabular}{c c c c} 
\hline\hline 
$n_1$ & $n_2$ & $n_3$ & ${I_1}(n_1 , n_2 , n_3)$ \\ [0.5ex] 
\hline 
2 & 2 & 0 & $1\times PRD$ \\ 
2 &	4 &	0 & $0.66\times PRD$ \\
1 &	1 &	1 & $1\times PRD$ \\
1 &	1 &	3 & $0.66\times PRD$ \\
3 &	1 &	3 & $0.45\times PRD$ \\
3 &	3 &	3 & $0.34\times PRD$ \\
2 &	2 &	2 & $0.5\times PRD$ \\
2 &	2 &	4 & $0.4\times PRD$ \\
1 &	1 &	5 & $0.55\times PRD$ \\ [1ex]

\hline 
\end{tabular}
\label{table_I1_D} 
\end{table}

\begin{table}[ht]
\caption{Integral results in the form of $I_2$ (equation \eqref{I2_D}) } 
\centering 
\begin{tabular}{c c c c} 
\hline\hline 
$n_1$ & $n_2$ & $n_3$ & ${I_2}(n_1 , n_2 , n_3)$ \\ [0.5ex] 
\hline 
1 &	1 &	1 &	$1\times PRD$ \\
1 &	1 &	3 &	$0.66\times PRD$ \\
1 &	1 &	5 &	$0.53\times PRD$  \\
1 &	3 &	1 &	$0.67\times PRD$  \\
1 &	3 &	3 &	$0.45\times PRD$  \\
2 &	2 &	2 &	$0.5\times PRD$   \\
1 &	3 &	3 &	$0.45\times PRD$  \\
2 &	2 &	2 &	$0.5\times PRD$   \\
2 &	2 &	4 &	$0.394\times PRD$ \\
1 &	5 &	1 &	$0.55\times PRD$  \\
2 &	4 &	2 &	$0.394\times PRD$ \\
3 &	3 &	1 &	$0.45\times PRD$  \\
3 &	3 &	3 &	$0.343\times PRD$ \\
3 &	5 &	1 &	$0.37\times PRD$  \\
4 &	4 &	2 &	$0.317\times PRD$  \\
5 &	5 &	1 &	$0.3043\times PRD$  \\
0 &	0 &	6 &	$0.55\times PRD^2$  \\
0 &	0 &	4 &	$0.667\times PRD^2$ \\
0 &	0 &	2 &	$1\times PRD^2$   \\
6 &	0 &	0 &	$0.55\times PRD^2$  \\
4 &	0 &	0 &	$0.667\times PRD^2$  \\
2 &	0 &	0 &	$1\times PRD^2$ \\ [1ex]

\hline 
\end{tabular}
\label{table_I2_D} 
\end{table}

\begin{table}[H]
\caption{Integral results in the form of $I_3$ (equation \eqref{I3_D}) } 
\centering 
\begin{tabular}{c c c c} 
\hline\hline 
$n_1$ & $n_2$  & ${I_3}(n_1 , n_2 )$ \\ [0.5ex] 
\hline 
6 & 6 & $0.3025\times PRD$ \\ 
6 &	4 & $0.367\times PRD$ \\
6 &	2 & $0.55\times PRD$ \\
4 &	6 & $0.367\times PRD$ \\
4 &	4 & $0.445\times PRD$ \\
4 &	2 & $0.667\times PRD$ \\
2 &	6 & $0.55\times PRD$ \\
2 &	4 & $0.667\times PRD$ \\
2 &	2 & $1\times PRD$ \\ [1ex]

\hline 
\end{tabular}
\label{table_I3_D} 
\end{table}

\begin{table}[H]
\caption{Integral results in the form of $I_4$ (equation \eqref{I4_D}) } 
\centering 
\begin{tabular}{c c c c} 
\hline\hline 
$n_1$ & $n_2$ & ${I_4}(n_1 , n_2 )$ \\ [0.5ex] 
\hline 
4 & 2 & $0.66 \times PRD$ \\ 
2 &	4 & $0.64 \times PRD$ \\
2 &	2 & $1 \times PRD$ \\ [1ex]

\hline 
\end{tabular}
\label{table_I4_D} 
\end{table}

Other than terms classified into cases A-Z, a large amount of terms that is not possible to show due its volume remain. Indeed, these terms are all the functions of $u$, $u^{'}$ and $u^{''}$ that are concentrated around zero. They have considerable value only for $u\in [-1,1]$, $u^{'}\in [-1,1]$ and $u^{''}\in [-1,1]$. For these terms and due to assuming $PRD>>1$ in the case of our problem, the triple integral in \eqref{ebarate_integralie_mom3_D} can be calculated with keeping $P_r$ and $\sigma_0$ as symbolic variables and numerical integration with respect to $u$, $u^{'}$ and $u^{''}$. While the numerical integration is performed for three different values of $PRD$s namely, $PRD=15$, $PRD=20$ and $PRD=25$, the outcomes are nearly the same.
Also in cases A-Z mentioned above, and for calculating triple integral in \eqref{ebarate_integralie_mom3_D}, four integral forms are required that have been expressed as follows:
\begin{equation}
\label{I1_D}
I_1\left(n_1,n_2,n_3\right)=\int_{-\ \frac{PRD}{2}}^{+\frac{PRD}{2}}\int_{-\
\frac{PRD}{2}}^{+\frac{PRD}{2}}\int_{-\
\frac{PRD}{2}}^{+\frac{PRD}{2}}R_i^{n_1}R_j^{n_2}R_k^{n_3}dudu^{'}du^{''}\ \ \ \ ;
i\neq j,\ i\neq k,\ j\neq k
\end{equation} 
\begin{equation}
\label{I2_D}
I_2\left(n_1,n_2,n_3\right)=\int_{-\ \frac{PRD}{2}}^{+\frac{PRD}{2}}\int_{-\
\frac{PRD}{2}}^{+\frac{PRD}{2}}\int_{-\
\frac{PRD}{2}}^{+\frac{PRD}{2}}a_i^{n_1}a_j^{n_2}R_{i+j-2}^{n_3}dudu^{'}du^{''}\
\ \ i\neq j
\end{equation} 
\begin{equation}
\label{I3_D}
I_3\left(n_1,n_2\right)=\int_{-\ \frac{PRD}{2}}^{+\frac{PRD}{2}}\int_{-\
\frac{PRD}{2}}^{+\frac{PRD}{2}}\int_{-\
\frac{PRD}{2}}^{+\frac{PRD}{2}}a_i^{n_1}R_{(4-i)}^{n_2}\ \ dudu^{'}du^{''}
\end{equation} 
\begin{align}
\label{I4_D}
&{{\ I}_4\left(n_1,n_2\right)=\int_{-\ \frac{PRD}{2}}^{+\frac{PRD}{2}}\int_{-\
\frac{PRD}{2}}^{+\frac{PRD}{2}}\int_{-\
\frac{PRD}{2}}^{+\frac{PRD}{2}}a_i^{n_1}R_i^{n_2}dudu^{'}du^{''}=} \\ \nonumber
&{\ \ \ \ \ \ \ \ \ \ \ \ \ \ \ \int_{-\
\frac{PRD}{2}}^{+\frac{PRD}{2}}\int_{-\ \frac{PRD}{2}}^{+\frac{PRD}{2}}\int_{-\
\frac{PRD}{2}}^{+\frac{PRD}{2}}a_j^{n_1}R_{j+1}^{n_2}dudu^{'}du^{''}=} \\ \nonumber
&{\ \ \ \ \ \ \int_{-\
\frac{PRD}{2}}^{+\frac{PRD}{2}}\int_{-\ \frac{PRD}{2}}^{+\frac{PRD}{2}}\int_{-\
\frac{PRD}{2}}^{+\frac{PRD}{2}}a_k^{n_1}R_{k-1}^{n_2}dudu^{'}du^{''} \ ;\
i\in{}\left\{1,3\right\}\ ,\ j\in{}\left\{1,2\right\},k\in{}\left\{2,3\right\}}
\end{align}
Where in \eqref{I1_D}, \eqref{I2_D} and \eqref{I3_D}, $i,j,k\in{}\left\{1,2,3\right\}$. These integrals are calculated in the tables \ref{table_I1_D}, \ref{table_I2_D}, \ref{table_I3_D} and \ref{table_I4_D}.
Finally summing all related terms, the third order moment can be expressed as follows:

\begin{equation}
\label{mom3_b_1_final_D}
{\mu{}}_3^{(1)}=\left\{\begin{array}{
cccc}
R^3k^3{\Gamma{}}^6\times{}\left(110592{\sigma{}}_0^{18}\right)+ \\
\frac{R^3k^3{\Gamma{}}^6}{PRD}\times{}\left(\begin{array}{l}5.16\times{}{10}^6{\sigma{}}_0^{18}+4.977\times{}{10}^5 P_r{\sigma{}}_0^{16}+8.3\times{}{10}^4 P_r^2{\sigma{}}_0^{14}
\\
+3.8\times{}{10}^3 P_r^3{\sigma{}}_0^{12}\end{array}\right)+ \\
\frac{R^3k^3{\Gamma{}}^6}{PRD^2}\times{}\left(\begin{array}{l}1.0538\times{}{10}^8{\sigma{}}_0^{18}+2.308\times{}{10}^7 P_r{\sigma{}}_0^{16}+
\\
8.133\times{}{10}^6 P_r^2{\sigma{}}_0^{14}
+1.306\times{}{10}^6 P_r^3{\sigma{}}_0^{12}+ \\
9.956\times{}{10}^4 P_r^4{\sigma{}}_0^{10}+3.479\times{}{10}^3 P_r^5{\sigma{}}_0^8
\\
+43.56 P_r^6{\sigma{}}_0^6\end{array}\right)+ \\
\frac{R^3k^3{\Gamma{}}^6}{PRD^3}\times{}\left(\begin{array}{l}4.671\times{}{10}^8 P_r{\sigma{}}_0^{16}+3.241\times{}{10}^8  P_r^2{\sigma{}}_0^{14}+ \\
1.027\times{}{10}^8 P_r^3{\sigma{}}_0^{12}
+1.647\times{}{10}^7 P_r^4{\sigma{}}_0^{10}+ \\
1.451\times{}{10}^6 P_r^5{\sigma{}}_0^8+7.232\times{}{10}^4 P_r^6{\sigma{}}_0^6
\\
+2.014\times{}{10}^3 P_r^7{\sigma{}}_0^4+28.97 P_r^8{\sigma{}}_0^2+0.1664 P_r^9\end{array}\right)
\end{array}\right\}
\end{equation}
Replacing $P_r=0$ in equation \eqref{mom3_b_1_final_D} one would find:
\begin{equation}
\label{mom3_b_0_final_D}
{\mu{}}_3^{(0)}=\
\frac{R^3k^3{\Gamma{}}^6}{PRD^3}\times{}\left\{\begin{array}{l}PRD^3\times{}\left(110592{\sigma{}}_0^{18}\right)+PRD^2\times{}\left(5.16\times{}{10}^6{\sigma{}}_0^{18}\right)
\\
+PRD\times{}\left(1.0538\times{}{10}^8{\sigma{}}_0^{18}\right)\end{array}\right\}
\end{equation}

\bibliographystyle{IEEEtranTCOM}
\bibliography{IEEE_TCOM}
%




\end{document}